\documentclass[a4paper,10pt]{amsart}
\usepackage[utf8]{inputenc}
\usepackage[T1]{fontenc}
\usepackage{graphicx} 
\usepackage{longtable}
\usepackage{graphicx}
\usepackage{subcaption}
\usepackage{mathtools}
\usepackage{color}
\usepackage{enumitem}

\usepackage{cite}
\usepackage{url}
\usepackage{amsmath}
\usepackage{amssymb}
\usepackage{amsthm}
\usepackage{setspace}
\usepackage[english]{babel}
\usepackage{mathrsfs}
\usepackage{mathtools}
\usepackage{xcolor}
\usepackage{mathrsfs}
\usepackage{hyperref} 
\usepackage[dots]{12many}

\usepackage{tabularx}
\DeclareMathOperator{\sgn}{\mathrm{sgn}}

\newcommand{\dd}{\mathrm{d}}
\DeclareMathOperator{\rank}{\mathrm{rank}}

\newcommand{\N}{\mathbb{N}}
\newcommand{\C}{\mathbb{C}}
\newcommand{\R}{\mathbb{R}}

\newcommand{\iu}{\mathrm{i}} 
\newcommand{\HH}{\mathscr{H}}

\newcommand{\spectrum}{\sigma} 
\newcommand{\resolvent}{\rho} 

\newcommand{\set}[2]{\left\lbrace #1 \, \middle\vert \, #2 \right\rbrace}  
\newcommand{\sclrp}[2]{\left\langle #1 \middle| #2 \right\rangle} 
\newcommand{\cc}[1]{\overline{#1}}	

\newcommand{\vertices}{\mathcal{V}} 
\newcommand{\intes}{\mathcal{I}} 
\newcommand{\extes}{\mathcal{E}} 
\newcommand{\edges}{\mathcal{J}} 
\newcommand{\graph}{\mathbf{G}}
\newcommand{\gb}{\partial} 
\newcommand{\vertspace}{\mathscr{F}} 
\newcommand{\boundspace}{\mathscr{G}} 
\newcommand{\orient}{\rho} 

\newcommand{\iedgea}{a} 
\newcommand{\iedgeb}{b} 
\newcommand{\iedge}{\left(\iedgea ,\iedgeb \right)} 
\newcommand{\eedge}{I} 

\newcommand{\pauli}[1]{\sigma_{#1}} 
\newcommand{\diracop}{- \iu \partial \otimes \pauli{1} + m \otimes \pauli{3} } 

\newcommand{\vectorComp}[1]{\begin{pmatrix} #1 \end{pmatrix}} 
\newcommand{\generalVector}[3][]{\vectorComp{#2^1_{#3}#1 \\ #2^2_{#3}#1 } } 
\newcommand{\idk}{\alpha} 
\newcommand{\sink}[1]{\sin \left( k(z) #1 \right)} 
\newcommand{\cosk}[1]{\cos \left( k(z) #1 \right)} 

\newcommand{\bm}{\Gamma} 
\newcommand{\bma}{\Gamma^1} 
\newcommand{\bmb}{\Gamma^2} 
\newcommand{\refoperator}{D^0} 
\newcommand{\Kreinfun}{\mathscr{Q}} 

\newcommand{\theother}[1]{\tilde{#1}} 
\newcommand{\tobm}{\theother{\Gamma}} 
\newcommand{\tobma}{\theother{\Gamma}^1} 
\newcommand{\tobmb}{\theother{\Gamma}^2} 

\newtheorem{theorem}{Theorem}[section]
\newtheorem{lemma}[theorem]{Lemma}
\newtheorem{corollary}[theorem]{Corollary}
\newtheorem{proposition}[theorem]{Proposition}

\newtheorem{definition}[theorem]{Definition}
\theoremstyle{definition}
\newtheorem{remark}[theorem]{Remark}

\renewcommand{\phi}{\varphi} 
\renewcommand{\epsilon}{\varepsilon}
\renewcommand{\Im}{\mathrm{Im}}
\renewcommand{\Re}{\mathrm{Re}}
\newcommand{\ee}{\mathrm{e}} 

\DeclareMathOperator{\Dom}{Dom}
\DeclareMathOperator{\Ran}{Ran}
\DeclareMathOperator{\Ker}{Ker}

\title{Non-self-adjoint Dirac operators on graphs}

\date{April 8, 2025}

\begin{document}

\author[M. Holzmann]{Markus Holzmann}
\address{Institut f\"{u}r Angewandte Mathematik\\
Technische Universit\"{a}t Graz\\
 Steyrergasse 30, 8010 Graz, Austria\\
E-mail: {\tt holzmann@math.tugraz.at}}

\author[V. R\r{u}\v{z}ek]{V\'{a}clav R\r{u}\v{z}ek}
\address{Department of Mathematics, Faculty of Nuclear Sciences and Physical Engineering\\
Czech Technical University in Prague \\
Trojanova 13, 120 00, Prague, Czechia \\
E-mail: {\tt ruzekva2@fjfi.cvut.cz}}

\author[M. Tu\v{s}ek]{Mat\v{e}j Tu\v{s}ek}
\address{Department of Mathematics, Faculty of Nuclear Sciences and Physical Engineering\\
Czech Technical University in Prague \\
Trojanova 13, 120 00, Prague, Czechia \\
E-mail: {\tt matej.tusek@fjfi.cvut.cz}}

\begin{abstract}
In this paper we introduce and study generally non-self-adjoint realizations of the Dirac operator on an arbitrary finite metric graph. Employing the robust boundary triple framework, we derive, in particular, a variant of the Birman Schwinger principle for its eigenvalues, and with an example of a star shaped graph we show that the point spectrum may exhibit diverse behaviour. Subsequently, we find sufficient and necessary conditions on transmission conditions at the graph's vertices under which the Dirac operator on the graph is symmetric with respect to the parity, the time reversal, or the charge conjugation transformation.
\end{abstract}

\maketitle

\section{Introduction}

Schr\"{o}dinger operators on metric graphs represent a significant class of quantum Hamiltonians, and as such, they have been the subject of extensive scrutiny. They provide toy models, known as quantum graphs, to describe the motion  of a non-relativistic quantum particle in a system of wires of infinitesimal diameter, but they have been used also in various other situations when propagation of waves in an effectively one-dimensional medium is considered \cite{Al_1983, An_1981, AvRaZu_1988, Bu_1985, RuSc_1953}.  Mathematically, the medium is described by a metric graph, which is roughly speaking a collection of vertices connected by a network of edges, and the Hamiltonian acts as a  Schr\"{o}dinger operator on each of the edges and is made self-adjoint by choosing properly transmission (also called boundary or vertex) conditions at every vertex. The literature on the quantum graphs is extremely vast, we just refer to the standard monograph \cite{BerkolaikoKuchment} as a starting point.

In this paper we will be concerned with relativistic quantum graphs, i.e., instead of Schr\"{o}dinger operators we will consider Dirac operators on the edges, which act on  suitable spaces of $\C^2$-valued functions as 
$$-\iu\partial\otimes\pauli{1}+m\otimes\pauli{3},$$
where $m\geq 0$, $\partial$ denotes the first weak derivative, and $\pauli{1}, \pauli{3}$ are the usual Pauli matrices, see \eqref{def_Pauli_matrices}.
There are significantly less results in this setting and most of them appeared in the last fifteen years. In this time period there has been a massive new wave of interest in Dirac operators kindled by the extraction of graphene by Geim and Novoselov followed by the Nobel Prize in Physics in 2010. Although graphene itself is effectively described by the two-dimensional Dirac operator, the one dimensional operator is also physically relevant, as a proof-of-principle quantum simulation of the one-dimensional Dirac equation using a single trapped ion was performed in \cite{NatureLett_10}, cf. \cite{Le_etal_2013} for an experimental realization with $^{87}\text{Rb}$ Bose-Einstein condensates. Also note that the effective dynamics of ultracold atoms in bichromatic optical lattices is governed by the one-dimensional Dirac equation \cite{Wi_etal_2011}.

Similarly to the non-relativistic setting, the spectral properties of a Dirac operator on a graph are encoded both in the graph topology and the transmission conditions at the graph's vertices. The latter are typically set in a way that the operator on the graph is self-adjoint. This is achieved in several (sometimes equivalent) ways. The von Neumann theory of self-adjoint extensions of symmetric operators was used in the pilot paper \cite{BuTr_1990} on  the relativistic quantum graphs and also, e.g., in \cite{BoHa_2003,AdLaTrWi_16}, where the domains of the self-adjoint extensions were described elegantly as maximal isotropic subspaces with respect to a certain form -- a strategy adopted from \cite{KoSc_1999} dealing with the non-relativistic setting. Another approach, based on prescribing a unitary scattering matrix at every vertex, was applied in \cite{Ha_2008, Ha_2023}, where also a comparison with the previous method was provided. Finally, the notion of boundary triples, which can be seen as a modern reformulation of the von Neumann extension theory, was used in  \cite{BoCaTe_2021_proc,BoCaTe_2021,GeTr_2021,Po_2008}. For the spectral analysis for special choices or classes of graphs and/or transmission conditions we refer to \cite{BuTr_1990, BaRaRa_2022, BoCaTe_2021_proc,AdLaTrWi_16}.

The first part of the current paper is devoted to a construction of closed but not necessarily self-adjoint realizations of the Dirac operator on any finite metric graph. To this aim, we will rely on the boundary triple framework, which is sufficiently robust to deal with the self-adjoint and non-self-adjoint realizations in a uniform manner.
Boundary triples and their induced Krein $\Kreinfun$- or Weyl functions are a tool from extension theory that allow to describe all closed intermediate extensions of a given closed and symmetric operator $S$. Such a boundary triple always consists of an additional Hilbert space $\boundspace$ and boundary maps $\bma, \bmb: \Dom S^* \rightarrow \boundspace$, such that an abstract version of Green's identity is fulfilled and $(\bma, \bmb)$ is surjective, see Section~\ref{sec:BT} for details. For our application to Dirac operators on finite graphs, we can choose $\boundspace \equiv \C^N$ with an appropriate $N < +\infty$. We will focus on those intermediate extensions of $S$ that are described via the abstract boundary conditions
\begin{equation} \label{eq:AB_TC_intro}
A\bma\Phi=B\bmb\Phi
\end{equation}
for bounded operators (in fact, $N\times N$ matrices) $A,B$ in $\boundspace$. Note that with any boundary triple there comes a reference operator $S^0$, which is the restriction of $S^*$ to those functions satisfying $\bma f = 0$, from which the induced Krein $\gamma$-field and the induced Krein $\Kreinfun$- or Weyl function $\Kreinfun(\cdot)$ are constructed. Information about the spectra of the closed extensions with the boundary conditions~\eqref{eq:AB_TC_intro} is then encoded in $A - B \Kreinfun(z)$ with $z$ belonging to the resolvent set of $S^0$.

In our situation of Dirac operators on finite metric graphs, we will construct a boundary triple as a direct sum of boundary triples for the individual edges.  A similar strategy was used in \cite{BoCaTe_2021_proc,BoCaTe_2021,GeTr_2021} but also earlier for the Dirac operator with $\delta$-interactions supported on a sequence of points \cite{CaMaPo_2013}. We choose the values of the boundary maps $\bma \Phi, \bmb \Phi$ as suitably weighted evaluations of the first and second component of the $\C^2$-valued ``spinors'' $\Phi$ belonging to the domain of the maximal Dirac operator on the graph at the endpoints of the edges. Here, one has to treat finite/internal and infinite/external edges differently. We use essentially the same boundary triple for the internal edges as it was used in \cite{GeTr_2021} but different from the triple used in the other cited papers.  It has the advantage of inducing a reference operator that is described by the Dirichlet boundary conditions for the first components of the functions in the operator domain. This means that both end-points of the edge are treated in the same way. By construction, the total reference operator is given by the direct sum of the reference operators on all edges and thus its spectrum and resolvent can be computed explicitly. Note that also the triple derived in \cite{Po_2008} from the so-called first order boundary triple introduced in \cite{Po_2007} yields the same reference operator on the whole graph.

Having the boundary triple, the  reference operator, the induced Krein $\gamma$-field  and the induced Krein $ \Kreinfun$-function in hand, it is rather immediate to write down a Birman Schwinger principle for the eigenvalues and a Krein like formula for the resolvent. From the physical perspective, local transmission conditions, that do not mix boundary values at different vertices, are favoured. As usual in the literature on Dirac operators on metric graphs, we write the vertex conditions in the form~\eqref{eq:AB_TC_intro}, where $A,B$ are (in the case of a finite graph) rectangular matrices. The boundary evaluations $\bma\Phi$ and $\bmb\Phi$ are given by direct sums with respect to the ``edge space'' that is just the direct sum of the boundary spaces for individual edges. On the other hand, for \eqref{eq:AB_TC_intro} being local, we require $A,B$ to  have a block matrix structure with respect to the ``vertex space'' that collects the boundary values vertex by vertex. To overcome this inconsistency, we construct (via an explicit algorithm)  a linear mapping $W$ that will send one of these spaces to the other and instead of \eqref{eq:AB_TC_intro} we will work with modified conditions
\begin{equation*}
A W\bma\Phi=B W\bmb\Phi.
\end{equation*}
The Dirac operator on a graph with these transmission conditions will be denoted by $D^{\Lambda_{A,B}}$, see Definition \ref{def:Dirac_on_graph}. While in many results our focus will be on local transmission conditions, we note that most of them remain true also for non-local transmission conditions. As described above, the boundary triple machinery allows to analyse the spectrum of $D^{\Lambda_{A,B}}$, see Theorem~\ref{theo:spectrum_AB} for an abstract result, and in Section \ref{sec:star_shaped} we show the applicability of the result with a fundamental example of a star shaped graph with infinite edges.

The second part of our paper is devoted to the study of symmetries of a generally non-self-adjoint Dirac operator on a graph. We were partially motivated by the paper \cite{HuKrSi_2015}, where generally non-self-adjoint transmission conditions are classified for usual (non-relativistic) quantum graphs, but instead of the operator classes our classification is motivated by physical symmetries. Of course, these symmetries have immediate spectral consequences. As far as we know, only very few papers deal with non-self-adjoint transmission conditions in the relativistic setting. In \cite{KrMuNi_2021} and the follow-up paper \cite{KrMuNi2_2021}, well-posedness of the time dependent Dirac equation  on a graph with rather general transmission conditions is discussed, and in \cite{Ri_2024} ``the scalar Dirac operator'', i.e., just the momentum operator, and very special generally non-self-adjoint transmission conditions, which make  it possible to link the spectrum to the appearance of cycles in the graph, are considered.  

First, we look at a parity operator. The transform $\psi(x)\mapsto\psi(-x)$ induced by the space inversion does not commute with the Dirac operator, as the latter is a first order differential operator, and so the orientation matters, cf. \cite{Ca_99} or \cite[Section 2.2]{BerkolaikoKuchment}. Mimicking established considerations in three spatial dimensions, cf. \cite{thaller}, we will derive that the mapping $P$ defined edge-wise as $(P\psi)(x):=\sigma_3\psi(-x)$ is a good candidate for the one-dimensional parity operator. Let us emphasise that strictly speaking $P$ maps functions on an edge to functions on the edge with opposite orientation (that will be, by our convention, parametrized by the inverted coordinate).  We will show that the Dirac operator $D^{\Lambda_{A,B}}$ with arbitrary choice of $A,B$ obeys $\tilde{D}^{\Lambda_{A,B}}=PD^{\Lambda_{A,B}}P^{-1}$, where $\tilde{D}^{\Lambda_{A,B}}$ stands for the Dirac operator with the same transmission conditions at all vertices but with opposite orientations of the edges where $P$ was applied; see Theorem \ref{thmPDLambda} for a precise formulation. In this sense, the change of orientation does not matter when combined with $\sigma_3$-transform of the ``spinor'' components.

Similarly, we will arrive at the time reversal transformation which will act edge-wise as $(T\psi)(x):=\sigma_3\overline{\psi(x)}$. This corresponds to a one-dimensional analogue of the time reversal transformation in three spatial dimensions, cf. \cite[Equation (3.159)]{thaller}.  Note that a different candidate for this transformation on a graph was considered in \cite{BoHa_2003, Ha_2008}. It does not commute with the Dirac operator on a line unless $m=0$ and when considered on a graph it commutes with the Dirac operator only for very special choices of transmission conditions and graph topologies. To get less restrictive conditions for the time reversal symmetry, the authors of \cite{BoHa_2003} suggest considering only graphs with paired edges of opposite orientations, which is effectively equivalent to introducing four-component spinors.  With our choice, the time reversal transformation always commutes with the Dirac operator on a line and it does also commute, e.g., with any $D^{\Lambda_{A,B}}$ such that $A$ and $B$ posses only real entries. See Corollary \ref{corollary_spectrum_time_reversal} for a sufficient and necessary condition for $D^{\Lambda_{A,B}} T=T D^{\Lambda_{A,B}}$.

The last symmetry transformation studied in this paper is the charge conjugation. Unlike the two previous transformations, it does not stem from the invariance under Poincar\'{e} transformations but it is rather an internal symmetry connecting the negative energy subspace of particles with the positive energy subspace of antiparticles. Following the same argument as in \cite[Section 1.4.6]{thaller}, we will derive that on every edge it acts as $(C\psi)(x):=\sigma_1\overline{\psi(x)}$. A necessary and sufficient condition for $D^{\Lambda_{A,B}}$ to commute with the charge conjugation is presented in Corollary \ref{corollary_C}. Although it may seem a bit complicated, it simplifies considerably for the case of a star shaped graph with external outgoing edges only, when it is satisfied, e.g., for any pair $A,B$ such that $A=\cc{B}$; cf. Remark \ref{rem:C_symm_star}.

The paper is organized as follows. Section \ref{sec:prelim} is of preliminary nature. After collecting some basic notation in Section \ref{sec:notation}, we introduce further notation concerning metric graphs and function spaces on them in Section \ref{sec:graphs}. In Section \ref{sec:BT}, an overview of the boundary triple framework is presented, while the  proof of a variant of the Birman Schwinger principle and the Krein resolvent formula is relocated to Appendix \ref{appendix_relations}. Dirac operators on graphs are introduced and their basic spectral properties are derived in Section \ref{sec:operators}. First, the Dirac operators are considered just on internal and external edges in Sections \ref{secIedge} and \ref{secEedge}, respectively. In Section \ref{sec:graph}, we put all previous results together to construct the Dirac operator on a finite metric graph. The spectral results deduced in a general setting are then demonstrated with an important example of a star shaped graph in Section \ref{sec:star_shaped}. Section \ref{sec:symm} is devoted to the study of various symmetries. After deriving their form from basic principles, we look at a sort of parity transformation in Section \ref{sec:parity}. Then we investigate the time reversal transformation and the charge conjugation in Sections \ref{sec:Treversal} and \ref{sec:Ctrafo}, respectively.

\section{Preliminaries} \label{sec:prelim}
In this section, we will first introduce some basic notation. Then we will present further notation, terminology, and conventions related to the metric graphs. Finally, we will recall basic definitions and results in the boundary triple framework. Some of these results will be reformulated in the way that suits our further purposes. 

\subsection{Notations} \label{sec:notation}
Let us stress that all vector spaces are always considered over $\C$ and that the inner products in Hilbert spaces are assumed to be linear and antilinear in the second and the first argument, respectively. We will denote by $L^2(I;\C^2)$ the Hilbert space of (equivalence classes of a.e. identical) square-integrable $\C^2$-valued functions with respect to the Lebesgue measure on an interval $I\subset\R$. The Sobolev space $H^1(I;\C^2)$ consists of those elements of $L^2(I;\C^2)$ whose both components have weak derivatives that belong to $L^2(I)$. The functions in its subspace $H^1_0(I;\C^2)$ have the extra property that the traces of their both components vanish at the endpoints of $I$. When convenient we will identify $L^2(I;\C^2)$ with $L^2(I)\otimes\C^2$ or $L^2(I) \oplus L^2(I)$ and similarly for its subspaces.

We will use the usual notation for the Pauli matrices,
\begin{equation} \label{def_Pauli_matrices}
  \sigma_1 := \begin{pmatrix} 0 & 1 \\ 1 & 0 \end{pmatrix}, \qquad \sigma_2 := \begin{pmatrix} 0 & -\iu \\ \iu & 0 \end{pmatrix}, \qquad \sigma_3 := \begin{pmatrix} 1 & 0 \\ 0 & -1 \end{pmatrix},
\end{equation}
and we will denote the $2\times 2$ identity matrix by $\sigma_0$. The symbols $\sigma(A)$ and $\rho(A)$ will stand for the spectrum and the resolvent set of a linear operator $A$. By $\sigma_p(A),\sigma_c(A)$, and $\sigma_r(A)$ we will mean the point spectrum, the continuous spectrum, and the residual spectrum of $A$, respectively. The discrete spectrum $\sigma_d(A)$ of $A$ will be defined as the set of all isolated (in the spectrum) eigenvalues of $A$ of finite algebraic multiplicity.  The essential spectrum $\sigma_{ess}(A)$ of $A$ will be understood as its complement in the whole spectrum of $A$. Finally, $\C^+$ and $\C^-$ will stand for the upper and the lower open complex half-plane, respectively.

\subsection{Metric graphs} \label{sec:graphs}
In our setting, a graph $\graph$ consists of a non-empty finite set of vertices $\vertices$, a finite set of bounded/internal edges $\intes$, and a finite set of unbounded/external edges $\extes$.  The set of all edges is denoted by $\edges = \extes \cup \intes$ and it is assumed to be non-empty. The structure of the graph is imposed by a graph boundary map $\gb: \edges \to \left( \vertices \cup \lbrace\infty \rbrace \right) \times \left( \vertices \cup \lbrace \infty \rbrace \right)$ which assigns to every edge its endpoints. The element $\infty$ is reserved for the external edges to denote their ``end point at $\infty$'' but it is not considered as a vertex, i.e., we have
\begin{equation*}
\gb(\intes) \subset \vertices \times \vertices
\qquad
\gb(\extes) \subset \left( \vertices \times \lbrace \infty \rbrace \right) \cup \left( \lbrace \infty \rbrace \times \vertices \right).
\end{equation*}
The maps  $\gb_-$ and  $\gb_+ : \edges \to \vertices \cup \lbrace\infty\rbrace$, that are the components of the graph boundary map $\gb=(\gb_-,\gb_+)$, assign the initial and the terminal vertex (or $\infty$), respectively, to every edge, and therefore, they induce  an \emph{orientation} on it.

The set of all incident edges to a vertex $v\in\vertices$ is denoted by
\begin{equation*} 
\intes_v := \set{i\in\intes}{ v \in \gb i}, \quad
\extes_v := \set{e\in\extes}{v\in\gb e}, \quad
\edges_v := \intes_v \cup \extes_v.
\end{equation*}
The degree of the vertex $v\in\vertices$ is then defined as the number of incident edges,
\begin{equation*}
\deg v := \left\vert \edges_v \right\vert.
\end{equation*}
Eventually, we will make use of the vertex space $\vertspace_v$ for the vertex $v\in\vertices$ and the vertex space for the whole graph $\graph$ as the direct sum over all vertices,
\begin{equation} \label{eq:Fdecomp}
\vertspace_v := \C^{\deg v} \quad \text{and} \quad \vertspace := \bigoplus_{v\in\vertices} \vertspace_v.
\end{equation}

Mathematically, the structure of the oriented  graph $\graph = \left( \vertices, \edges, \gb \right)$ can be conveniently encoded in the associated incidence matrix. Let $\gb_{-}, \gb_{+}$ be the components of the graph boundary map $\gb$.
Then the incident matrix is the map $G: \left( \vertices \times \edges \right) \to \lbrace -1, 0, 1 \rbrace$ defined by,
\begin{equation} \label{def_incidence_matrix}
\forall v \in \vertices, \forall j \in \edges:
\qquad
G_{v,j} = 
\begin{cases}
1 & v = \gb_+ j, \\
-1 & v = \gb_- j, \\
0 & v \notin \gb j.
\end{cases}
\end{equation}
We emphasize that, in the current paper, multiple edges with the same endpoints, even in the same order, are allowed. On the other hand, loops are prohibited. However, regarding our aims, the latter is not a restriction at all since we can always insert an extra vertex on a loop and equip it with continuous transmission conditions that don't change the Dirac operator on the loop.

To turn the graph into a metric graph we endow it with a parametrization which associates each internal edge with a finite interval and each external edge with a half-line. We will adopt the following convention that respects the chosen orientation of the edges. Every internal edge $i\in\intes$ will be associated with an interval $I_i = \left( a_i, b_i \right)$, $0<b_i-a_i<+\infty$, in the way that $a_i$ and $b_i$ correspond to the initial and the terminal vertex of $i$, respectively. Every external edge $e\in\extes$ will be associated either with $I_e = (a_e,+\infty)$ or $I_e = (-\infty,b_e)$, $a_e,b_e\in\R$. We will apply the former choice if $\gb_- e\in\vertices$ and the latter choice whenever $\gb_+ e\in\vertices$. To deal later with both options in a uniform manner, we will introduce the following \emph{orientation map} $\orient: \extes \to \lbrace -1, 1 \rbrace$,
\begin{equation*}
\forall e \in \extes:
\qquad
\orient(e)
=
\begin{cases}
-1, & \gb_{-}e \in \vertices, \\
1, & \gb_{+}e \in \vertices.
\end{cases}
\end{equation*}
For convenience, we will sometimes write $\gb_{-1}$ and $\gb_{1}$ for $\gb_{-}$ and $\gb_{+}$, respectively. Then $\gb_{\orient(e)}e$ is always the vertex to which an external edge $e$ is attached.

Finally, we describe how an oriented graph $\graph = \left( \vertices, \edges, \gb \right)$ can be endowed with Hilbert spaces. Since the Dirac operator acts on $\C^2$-valued functions, it is natural to work in the Hilbert space
\begin{equation} \label{def_L_2_graph}
\HH = L^2( \graph; \C^2 ):=
\bigoplus_{j\in\edges} L^2 (I_j ; \C^2 ).
\end{equation}
Elements in $\HH$ will be denoted by $\Phi = (\phi_j)_{j \in \edges}$ with $\phi_j \in L^2(I_j; \C^2)$.
In a similar fashion, also Sobolev spaces on $\graph$ are defined by
\begin{equation} \label{def_Sobolev_space_graph}
\begin{split}
\widetilde{H}^1(\graph; \C^2) := \bigoplus_{j\in\edges} H^1 ( I_j ; \C^2 ), \\
\widetilde{H}^1_0 (\graph; \C^2 ) := \bigoplus_{j\in\edges} H^1_0 ( I_j ; \C^2 ).
\end{split}
\end{equation}
Note that elements in $\widetilde{H}^1(\graph; \C^2)$ are not necessarily continuous at the vertices $v \in \vertices$ and that elements in $\widetilde{H}^1_0 (\graph; \C^2 )$ have the value $0$ at all vertices $v \in \vertices$.
To evaluate functions $\Phi = (\phi_j)_{j \in \edges} \in \widetilde{H}^1( \graph; \C^2 )$ with $\phi_j\in H^1( ( a_j, b_j ), \C^2 )$ at vertices we write, assuming that $v\in\gb j \cap \vertices$,
\begin{equation*}
\phi_j(v)=
\begin{cases}
\phi_j(a_j) & v = \gb_- j, \\
\phi_j(b_j) & v = \gb_+ j.
\end{cases}
\end{equation*}

\subsection{Boundary triples} \label{sec:BT}

In this section, we first recall basic notions concerning the boundary triple framework. For further reading and proofs we refer to the monograph \cite{BeHaSn_2020}, the overview paper \cite{BrGePaBT_2008}, the textbook \cite{Schmudgen}, and the references therein. We will always assume that $S$ is a closed symmetric operator in a Hilbert space $\HH$ with the inner product $\langle\cdot|\cdot\rangle$ and $S^*$ is its adjoint.

\begin{definition} \label{def:BT}
Let $\boundspace$ be another Hilbert space with the inner product $\langle\cdot | \cdot\rangle_\boundspace$ and $\bma,\bmb:\, \Dom{S^*}\to\boundspace$ be linear mappings that satisfy the abstract \emph{Green identity}
\begin{equation} \label{eq:Green}
\langle\phi|S^*\psi\rangle-\langle S^*\phi|\psi\rangle=\langle\bma\phi|\bmb\psi\rangle_\boundspace-\langle\bmb\phi|\bma\psi\rangle_\boundspace \quad (\forall\phi,\,\psi\in\Dom{S^*})
\end{equation}
together with the condition that $(\bma,\bmb):\, \Dom{S^*}\to\boundspace\oplus\boundspace$ is surjective. Then the triple $(\bma,\bmb,\boundspace)$ is called a boundary triple for $S^*$.
\end{definition}

It is well known that if $N := \dim \Ker(S^*- i) = \dim \Ker (S^* + i)$, then there exists a boundary triple for $S^*$ with $\dim{\boundspace}=N$, see, e.g, \cite[Proposition 14.5]{Schmudgen}. We will further assume that this condition is satisfied. Given a closed linear relation $\Lambda$ in $\boundspace$ (i.e. a closed linear subspace of $\boundspace \oplus \boundspace$), we define the operator $S^\Lambda$ as the  restriction of $S^*$ to
\begin{equation} \label{dom_S_Lambda}
\Dom{S^\Lambda}:=\{\phi\in\Dom{S^*}\vert\, (\bma\phi,\bmb\phi)\in\Lambda\}.
\end{equation}
There is a one-to-one correspondence between all closed linear relations in $\boundspace$ and all closed intermediate extensions of $S$ (which are all closed operators $A$ satisfying $S \subset A \subset S^*$) given by $\Lambda\leftrightarrow S^\Lambda$. Many properties of $S^\Lambda$ are then encoded in the properties of $\Lambda$. For example, $S^\Lambda$ is self-adjoint in $\HH$ if and only if $\Lambda$ is self-adjoint in $\boundspace$, cf. \cite[Proposition 14.7]{Schmudgen}. In particular, this implies that $S^0:=S^{\{0\}\oplus\boundspace} = S^* \upharpoonright \Ker \bma$ is always self-adjoint. We will call it the \emph{reference operator}. For $z\in\rho(S^0)$, it is not difficult to show that the direct sum decomposition
\begin{equation*}
  \Dom S^* = \Dom S^0 \dot{+} \Ker(S^* - z) = \Ker \Gamma^1 \dot{+} N_z
\end{equation*}
with $N_z:=\Ker(S^*-z)$ holds. In particular, $\Gamma^1 \upharpoonright N_z$ is injective and one can introduce the so-called Krein $\gamma$-field at $z$ induced by the triple $(\bma,\bmb,\boundspace)$ by
\begin{equation} \label{eq:gammaField}
\gamma(z):=(\bma\upharpoonright N_z)^{-1},
\end{equation}
which turns out to be a bounded linear map from $\boundspace$ to $\HH$. To study the spectrum of $S^\Lambda$, it is also useful to introduce the following bounded linear operator in $\boundspace$,
\begin{equation*}
\Kreinfun(z):=\bmb\gamma(z),
\end{equation*}
which is called the Krein $\Kreinfun$-function (or often also the Weyl function) at $z$ induced by the triple $(\bma,\bmb,\boundspace)$.

Next, let us focus on the setting of the current paper when $\boundspace$ is finite-dimensional and  
\begin{equation} \label{def_Lambda}
  \Lambda = \bigl\{ ( f, f') \in \boundspace \oplus \boundspace\,|\, A f = B f' \bigr\},
\end{equation}
where $A, B$ are bounded and everywhere defined linear operators in $\boundspace$. The representation in~\eqref{def_Lambda} is particularly convenient for us, as then the extension $S^\Lambda$ given by~\eqref{dom_S_Lambda} is the restriction of $S^*$ onto those $f \in \Dom S^*$ that satisfy the abstract boundary conditions
\begin{equation*}
  A \bma f = B \bmb f.
\end{equation*}
Note that the choice of $A$ and $B$ is unique up to left-multiplication by a bijective operator in the following sense, cf. \cite[Proposition~1.10.4]{BeHaSn_2020}. 

\begin{proposition} \label{prop:equivRep}
Let $\Lambda$ be given by~\eqref{def_Lambda}, $\tilde\Lambda:=\{ ( f, f')\,|\, \tilde A f =\tilde B f' \}$  with a pair of bounded linear mappings $\tilde A, \tilde B$ in $\boundspace$, $\dim{\boundspace}\leq+\infty$, and
$$\mathscr{F}:=\mathrm{span}\{\Ran(A),\Ran(B)\},\quad \tilde{\mathscr{F}}:=\mathrm{span}\{\Ran(\tilde A),\Ran(\tilde B)\}.$$
Then $\Lambda=\tilde \Lambda$ if and only if there exists linear bijection $X:\mathscr{F}\to\tilde{\mathscr{F}}$ such that $\tilde A=X A$ and $\tilde B=X B$. 
\end{proposition}
\noindent
Let us emphasise that when $\dim{\boundspace}<+\infty$, one can always extend $X$ to a linear bijection on the whole space $\boundspace$, but necessarily $\dim(\mathscr{F})=\dim(\tilde{\mathscr{F}})$. 

Self-adjointness of $S^\Lambda$ can be characterized in terms of $A$ and $B$ from \eqref{def_Lambda}. 
\begin{proposition}[{\cite[Corollary~1.6]{BrGePaBT_2008}}] \label{proposition_Lambda_self_adjoint}
Let $\dim{\boundspace}<+\infty$ and $\Lambda$ be given by \eqref{def_Lambda}. Then $\Lambda$ (and, therefore, also $S^\Lambda$) is self-adjoint if and only if $AB^*=BA^*$ and the block matrix  $(A\vert B)$ has maximal rank.
\end{proposition}
\noindent
Note that the latter condition is satisfied if and only if $\det(AA^*+BB^*)\neq 0$.

In the next theorem we formulate a variant of the Birman Schwinger principle and the Krein resolvent formula for $S^\Lambda$ that will be useful in our application to Dirac operators on finite graphs. It may be viewed as a counterpart of \cite[Corollary~2.6.3]{BeHaSn_2020}, where the so-called parametric representation of relations is considered. 
While the following result is known in the self-adjoint setting, see, e.g., \cite[Theorem 1.33]{BrGePaBT_2008}, we are not aware of a formulation in the general non-self-adjoint setting. Hence, we provide its proof, for which a deeper analysis of the involved relations is necessary and which is, in many aspects, also true for infinite dimensional spaces $\boundspace$, in Appendix~\ref{appendix_relations} (see  Theorem~\ref{theorem_krein}).

\begin{theorem} \label{theorem_Birman_Schwinger}
Let $(\bma, \bmb, \boundspace)$ be a boundary triple for $S^*$ such that $\dim{\boundspace}<+\infty$, let $S^0 = S^* \upharpoonright \Ker \bma$, and let $\gamma$ and $\Kreinfun$ be the induced Krein $\gamma$-field and the induced Krein $\Kreinfun$-function, respectively. Moreover, let $\Lambda$ be  a linear relation in $\boundspace$ of the form~\eqref{def_Lambda} and let $S^\Lambda$ be given by~\eqref{dom_S_Lambda}. Then, the following is true:
\begin{itemize}
  \item[\textup{(i)}] For $z \in \rho(S^0)$ one has that $0 \in \sigma(A - B \Kreinfun(z))$ if and only if $z \in \sigma_p(S^\Lambda)$.
  \item[\textup{(ii)}] If $\rank (A|B) < \dim \boundspace$, then $\sigma(S^\Lambda) = \C$.
  \item[\textup{(iii)}] For $z \in \rho(S^0)$ one has that $0 \notin \sigma(A - B \Kreinfun(z))$ if and only if $z \in \rho(S^\Lambda)$ and in this case
  \begin{equation*} 
(S^\Lambda - z)^{-1} = (S^0 - z)^{-1} + \gamma(z) \bigl( A - B \Kreinfun(z) \bigr)^{-1} B \gamma(\overline{z})^*.
\end{equation*}
\end{itemize}
\end{theorem}

Finally, we recall a construction of direct sums of boundary triples, which will be useful to introduce boundary triples on metric graphs. The proof of the following result is, for finite index sets, straightforward and left to the reader; we remark that the statement may  not be true for infinite index sets; cf. \cite{CaMaPo_2013, ExKoMaNe_2018, KoMa_2010}.

\begin{theorem} \label{theorem_direct_sum}
Assume that $\mathcal{J}$ is a finite index set, that $S_j$, $j \in \mathcal{J}$, is a closed symmetric operator in a Hilbert space $\HH_j$ with equal deficiency indices, and $(\bma_j,\bmb_j,\boundspace_j)$, $j \in \mathcal{J}$, is a boundary triple for $S_j^*$ with the induced Krein $\gamma$-field $\gamma_j$ and the induced Krein $\Kreinfun$-function $\Kreinfun_j$. Moreover, set 
\begin{equation*}
  \boundspace := \bigoplus_{j \in \mathcal{J}} \boundspace_j, \quad \bma := \bigoplus_{j \in \mathcal{J}} \bma_j, \quad \text{and} \quad \bmb := \bigoplus_{j \in \mathcal{J}} \bmb_j.
\end{equation*}
Then, $(\bma, \bmb, \boundspace)$ is a boundary triple for $\bigoplus_{j \in \mathcal{J}} S_j^*$ and the values of the induced Krein $\gamma$-field $\gamma$ and the induced Krein $\Kreinfun$-function $\Kreinfun$ are given by
\begin{equation*}
  \gamma(z) = \bigoplus_{j \in \mathcal{J}} \gamma_j(z) \quad \text{and} \quad \Kreinfun(z) = \bigoplus_{j \in \mathcal{J}} \Kreinfun_j(z) \qquad \big(\forall z \in \bigcap_{j \in \mathcal{J}} \rho(S^*_j \upharpoonright \Ker \bma_j)\big).
\end{equation*}
\end{theorem}

\section{Dirac operators on edges and graphs} \label{sec:operators}

In this section we construct a boundary triple that is suitable to introduce and study Dirac operators on a finite graph $\graph$ and collect basic results concerning their spectra and resolvents. For this purpose, we construct first boundary triples on the single edges, for internal edges in Section~\ref{secIedge} and for external edges in Section~\ref{secEedge}. In Section~\ref{sec:graph} the boundary triple on the finite graph is constructed as the direct sum of the boundary mappings and spaces on the edges. Finally, in Section~\ref{sec:star_shaped} we show the applicability of this approach to the important example of a star graph.

\subsection{Dirac operator on an internal edge}\label{secIedge}

Throughout this subsection, let $-\infty < \iedgea < \iedgeb < + \infty$  be fixed. First, we discuss several facts about Dirac operators in the Hilbert space $\HH =L^2\left(\iedge\right) \otimes \C^2\equiv L^2(\iedge;\C^2)$. Every $\phi \in \HH$ viewed as an element of $L^2(\iedge;\C^2)$ will be written as a pair of $\C$-valued functions $\phi = \generalVector{\phi}{}$. 

For $m\geq 0$ we define the minimal realization $D^{min}$ and the maximal realization $D^{max}$ of the Dirac operator in $\HH$ by
\begin{equation}\label{eqDefDiracMaxIedge}
D^{max} = \diracop, \quad  \Dom D^{max} = H^{1}_{} ( \iedge; \C^2  ),
\end{equation}
and
\begin{equation*} 
D^{min} = \diracop, \quad \Dom D^{min} = H^{1}_{0} (\iedge; \C^2 ),
\end{equation*}
respectively.
Note that the minimal Dirac operator is closed and symmetric, and the maximal Dirac operator is its adjoint 
$D^{max} = (D^{min})^*$. In the next proposition we introduce a boundary triple for $D^{max}$.

\begin{proposition}
Define the two linear maps $\bma, \bmb: H^1 (\iedge; \C^2 ) \to \C^2$ by
\begin{align}\label{eqDefBTIedge}
\bma \phi &= \vectorComp{ \phi^1(\iedgea) \\  \phi^1(\iedgeb) },
&
\bmb \phi &= \vectorComp{ \iu \phi^2(\iedgea) \\ - \iu \phi^2(\iedgeb) }.
\end{align}
Then the triple $( \bma, \bmb, \C^2 )$ is a boundary triple for $D^{max}$.
\end{proposition}
\begin{proof}
First, integration by parts yields the abstract Green identity \eqref{eq:Green}.
Next, to show the surjectivity of $(\bma, \bmb)$, consider the functions $f_{a}(x) = \frac{x-a}{b-a}$ and $f_{b}(x) = \frac{b-x}{b-a}$. Then, it is not difficult to see that
\begin{equation*}
  \vectorComp{f_b \\ 0}, \vectorComp{f_a \\ 0}, \vectorComp{0 \\ -\iu f_b}, \vectorComp{0 \\ \iu f_a} \in H^1 ( \iedge; \C^2)
\end{equation*}
are mapped by $( \bma, \bmb )$ to the standard basis in $\C^4$. This yields the surjectivity of $( \bma, \bmb )$. 
Finally, $D^{min}$ is densely defined, closed and symmetric with adjoint $(D^{min})^* = D^{max}$. Therefore, we conclude from Definition \ref{def:BT} that $( \bma, \bmb, \C^2 )$ is a boundary triple for $D^{max}$.
\end{proof}

\begin{remark}
Note that a different boundary triple for the Dirac operator on a line segment was introduced in \cite{CaMaPo_2013} when studying Dirac operators with point interactions on a discrete set. The same triple was also used later in \cite{BoCaTe_2021} to deal with Dirac operators on graphs with Kirchhoff-type vertex conditions. We prefer our choice because it leads to a reference operator with identical boundary conditions at both endpoints, see \eqref{eq:refOpdef}. 
\end{remark}

The next goal is to compute the Krein $\gamma$-field and the Krein $\Kreinfun$-function induced by the boundary triple $( \bma, \bmb, \C^2 )$. For this purpose, we compute first the defect subspaces for $D^{max}$. In the following, we will frequently use the notations
\begin{equation} \label{def_k_z}
  k(z) = \sqrt{z^2 - m^2} \quad \text{with} \quad \arg k(z) \in \left[ 0, \pi \right)
\end{equation}
and, for $z \in \C \setminus \{ -m,  m \}$,
\begin{equation} \label{def_alpha}
  \idk(z) = \frac{k(z)}{z-m}=\frac{z+m}{k(z)}.
\end{equation}

\begin{proposition} \label{proposition_N_z_internal_edge}
Let $D^{max}$ be defined by~\eqref{eqDefDiracMaxIedge}. Then, for $z \in \mathbb{C}$ one has 
\begin{equation*}
  N_{z}=\Ker \left( D^{max} - z \right) = \textup{span}\, \{ \psi_{z, 1}, \psi_{z,2} \},
\end{equation*}
where for $z \in \C \setminus \lbrace -m, m\rbrace$
\begin{align} \label{def_psi_z}
\psi_{z, 1} (x) &= \vectorComp{ \cosk{x}  \\ \iu (\idk(z))^{-1} \sink{x} },
& 
\psi_{z, 2} (x) &= \vectorComp{ \iu \idk(z) \sink{x} \\ \cosk{x} }, 
\end{align}
and  for $z = \pm m$
\begin{align} \label{def_psi_m}
\psi_{\pm m, 1} (x) &= \vectorComp{ 1 \\ \iu (\pm m - m)x },
&
\psi_{\pm m, 2} (x) &= \vectorComp{ \iu (\pm m +m)x \\ 1 }.
\end{align}
\end{proposition}
\begin{proof}
Let $z \in \mathbb{C}$ be fixed. By definition, the defect subspace $N_z$ consists of solutions $\phi \in \Dom D^{max}$ of 
\begin{equation*}
\left(\diracop - z \otimes \sigma_0 \right) \phi = 0,
\end{equation*}
which are given by
\begin{equation}\label{eqsolutionOfDefectSubspaces}
\phi(x) = \exp\left[ \left( \iu z \pauli{1} - m \pauli{2} \right) x \right] \generalVector{\omega}{}, \quad x \in \iedge,
\end{equation}
with $\omega^1, \omega^2 \in \mathbb{C}$. Since $(a,b)$ is bounded, each such solution belongs to $H^1((a,b);\C^2)$. For $k(z) \neq 0$ (and thus $z \neq \pm m$) one gets via a direct calculation that
\begin{equation*}
\phi(x) = 
\begin{pmatrix}
\cosk{x} & \iu \idk(z) \sink{x} \\
\iu (\idk(z))^{-1} \sink{x} & \cosk{x}
\end{pmatrix}
\generalVector{\omega}{ } = \omega^1 \psi_{z, 1}(x) + \omega^2 \psi_{z, 2}(x),
\end{equation*}
while, similarly, for $k(z) = 0$, which is equivalent to  $z = \pm m$, one obtains that
\begin{equation*}
\phi(x)=
\begin{pmatrix}
1 & \iu (\pm m +m)x \\
\iu (\pm m - m)x & 1
\end{pmatrix}
\generalVector{\omega}{ } = \omega^1 \psi_{\pm m, 1}(x) + \omega^2 \psi_{\pm m, 2}(x).
\end{equation*}
\end{proof}

In the following corollary we introduce an alternative basis for $N_z$ which will be useful in our further considerations.

\begin{corollary}
Let $z \in \mathbb{C}$ such that  $\alpha(z) \sin(k(z) (b-a)) \neq 0$. Then the functions
\begin{equation}\label{eqDefEtazIedge}
\begin{split}
\eta_{z,1}(x)&= 
\frac{1}{\sink{(b-a)}} 
\vectorComp{\sink{(b-x)} \\ \frac{\iu}{\idk(z)}\cosk{(b-x)}},
\\
\eta_{z,2}(x)&=
\frac{1}{\sink{(b-a)}} 
\vectorComp{\sink{(x-a)} \\ \frac{-\iu}{\idk(z)}\cosk{(x-a)}},
\end{split}
\end{equation}
are also a basis of $N_z = \Ker (D^{max} - z)$. Moreover, if $z=m \neq 0$, then the functions
\begin{equation}\label{eqDefEtamIedge}
\begin{split}
\eta_{m,1}(x)&=
\frac{1}{b-a}
\vectorComp{ b-x \\ \frac{\iu}{2m}},
\\
\eta_{m,2}(x)&=
\frac{1}{b-a}
\vectorComp{ x-a \\ \frac{-\iu}{2m}},
\end{split}
\end{equation}
are also a basis of $N_m = \Ker (D^{max} - m)$.
\end{corollary}
\begin{proof}
  Let $\psi_{z,1}, \psi_{z,2}$ be the basis of $N_z$ from~\eqref{def_psi_z}\&\eqref{def_psi_m}. For $\alpha(z) \sin(k(z) (b-a)) \neq 0$ one has
\begin{equation*}
  \eta_{z,1} = \frac{\sin(k(z) b)}{\sin(k(z)(b-a))} \psi_{z,1} + \frac{\iu \cos(k(z) b)}{\alpha(z) \sin(k(z) (b-a))} \psi_{z,2}
\end{equation*}
and
\begin{equation*}
  \eta_{z,2} = -\frac{\sin(k(z) a)}{\sin(k(z)(b-a))} \psi_{z,1} - \frac{\iu \cos(k(z) a)}{\alpha(z) \sin(k(z) (b-a))} \psi_{z,2}
\end{equation*}
and therefore, $\eta_{z,1}$, $\eta_{z,2}$ is a basis of $N_z$.
Likewise, as
\begin{equation*}
  \eta_{m,1} =\frac{b}{b-a}\psi_{m,1} +  \frac{\iu}{2m(b-a)}\psi_{m,2} \quad \text{and} \quad
  \eta_{m,2} = - \frac{a}{b-a} \psi_{m,1} -\frac{\iu}{2m(b-a)} \psi_{m,2},
\end{equation*}
the functions $\eta_{m,1}$ and $\eta_{m,2}$ form a basis of $N_m$.
\end{proof}

In the next proposition we compute the spectrum and the resolvent of the reference operator $D^0 := D^{max} \upharpoonright {\Ker \bma}$, i.e.,
\begin{equation} \label{eq:refOpdef}
D^0=\diracop, \quad  \Dom D^0 =\left\{\phi\in H^{1}_{} ( \iedge; \C^2  )\vert \, \phi^1(a)=\phi^1(b)=0\right\}.
\end{equation}
Recall that the operator $D^0$ is self-adjoint in $\HH$; cf. Section~\ref{sec:BT}. 

\begin{proposition} \label{proposition_D_0_internal_edge}
For $\refoperator$, we have:
\begin{itemize}
  \item[\textup{(i)}] $\sigma(D^0)$ is purely discrete and consists of simple eigenvalues,
\begin{equation}\label{eqSpectrumRefOpOnIedge}
\spectrum (\refoperator)
=
\spectrum_{p} (\refoperator)
=
\lbrace - m \rbrace \cup
\set{\pm \sqrt{m^2 + \frac{l^2 \pi^2}{(b-a)^2}}}{l \in \N}.
\end{equation}
\item[\textup{(ii)}] Define for $w\in\rho(D^0)\setminus\{m\}$ the function $R_w: (a, b) \times (a,b) \rightarrow \C^{2 \times 2}$ by
\begin{equation*}
  R_w(x,y) = \alpha(w)\sin(k(w) (b-a)) \begin{cases} \eta_{w,1}(x) \eta_{\overline{w},2}(y)^*, & x-y > 0, \\ \eta_{w,2}(x) \eta_{\overline{w},1}(y)^*, & x-y < 0, \end{cases}
\end{equation*}
and the function $R_m: (a, b) \times (a,b) \rightarrow \C^{2 \times 2}$ by
\begin{equation*}
  R_m(x,y) = 2 m (b-a) \begin{cases} \eta_{m,1}(x) \eta_{m,2}(y)^*, & x-y > 0, \\ \eta_{m,2}(x) \eta_{m,1}(y)^*, & x-y < 0. \end{cases}
\end{equation*}
Then for $z \in \rho(D^0)$ the resolvent of $D^0$ acts as
\begin{equation*}
(D^0 - z)^{-1} \phi(x)
=
\int_{\iedgea}^{\iedgeb} R_z(x,y) \phi(y) \dd y, \qquad (\forall\phi \in L^2((a,b); \C^2)).
\end{equation*}
\end{itemize}
\end{proposition}
As we will see in the proof, $z \in \rho(D^0)$ if and only if $\alpha(z) \sin(k(z) (b-a)) \neq 0$ or $z=m$, see~\eqref{equation_eigenvalues}. Therefore, the functions $\eta_{z,1}$ and $\eta_{z,2}$ defined in~\eqref{eqDefEtazIedge}\&\eqref{eqDefEtamIedge} and thus, also the integral kernel $R_z$, are well-defined for all $z \in \rho(D^0)$.
\begin{proof}[Proof of Proposition~\ref{proposition_D_0_internal_edge}]
First, we show the claim in~(i). As $D^0$ is self-adjoint and $\Dom D^0 \subset H^{1}_{} ( \iedge; \C^2  )$ is compactly embedded in $\HH = L^2(\iedge; \C^2)$, the spectrum of $D^0$ is purely discrete. Moreover, $z\in \sigma_p(D^0)$ if and only if there exists $0 \neq \varphi \in N_z$ such that $\bma \varphi = 0$. By Proposition~\ref{proposition_N_z_internal_edge} this is for $z \neq \pm m$ equivalent to the existence of $(\omega^1, \omega^2) \neq (0,0)$ such that
\begin{equation}\label{eqInvertibilityZIedge}
\bma (\omega^1 \psi_{z, 1} + \omega^2 \psi_{z, 2})
=
\begin{pmatrix}
\cosk{a} & \iu \idk(z) \sink{a} \\
\cosk{b} & \iu \idk(z) \sink{b} \\
\end{pmatrix}
\generalVector{\omega}{ }
=
\vectorComp{0 \\ 0},
\end{equation}
which is the case if and only if
\begin{equation} \label{equation_eigenvalues}
\det \begin{pmatrix}
\cosk{a} & \iu \idk(z) \sink{a} \\
\cosk{b} & \iu \idk(z) \sink{b} \\
\end{pmatrix} = \iu \idk(z) \sink{\left( b-a \right)} = 0.
\end{equation}
Therefore, all $z \in \mathbb{R}$ such that $k(z) (b-a) = l \pi$ for some $l \in \mathbb{Z}\setminus \lbrace 0 \rbrace$, i.e.
\begin{equation} \label{eigenalues1}
  z = \pm \sqrt{m^2 + \frac{l^2 \pi^2}{(b-a)^2}}, \qquad l \in \mathbb{N},
\end{equation}
are eigenvalues of $D^0$. Moreover, as the matrix in~\eqref{eqInvertibilityZIedge} is not identically zero, all of these eigenvalues have multiplicity one.

Similarly, by Proposition~\ref{proposition_N_z_internal_edge} one has for $z = \pm m$ that $0 \neq \varphi \in N_{\pm m}$ and $\bma \varphi = 0$ if and only if there exists $(\omega^1, \omega^2) \neq (0,0)$ such that
\begin{equation*}
\bma (\omega^1 \psi_{\pm m, 1} + \omega^2 \psi_{\pm m, 2})
=
\begin{pmatrix}
1 &  \iu \left( \pm m + m \right) a \\
1 &  \iu \left( \pm m + m \right) b \\
\end{pmatrix}
\generalVector{\omega}{}
=
\vectorComp{0\\0}.
\end{equation*}
From this, one sees, since $b-a>0$, for $m \neq 0$ that $z = m$ is not an eigenvalue. On the other hand $-m$ is always a simple eigenvalue. Together with~\eqref{eigenalues1}, we conclude that \eqref{eqSpectrumRefOpOnIedge} is true.

To show (ii), consider the unitary matrix
\begin{equation*}
  U = \begin{pmatrix} 1 & 0 \\ 0 & \iu \end{pmatrix}
\end{equation*}
and the operator $\widetilde{D}^0 := U^* D^0 U$. This operator is given explicitly by
\begin{equation*}
  \widetilde{D}^0 = \iu \partial \otimes \sigma_2 + m \otimes \sigma_3, \qquad \Dom \widetilde{D}^0 = \Dom D^0.
\end{equation*}
Moreover, consider for $z \in \rho(\widetilde{D}^0) = \rho(D^0)$ the functions 
\begin{equation*}
  u_a := U^* \eta_{z,2} \quad \text{and} \quad u_b := U^* \eta_{z,1},
\end{equation*}
where $\eta_{z,1}$ and $\eta_{z,2}$ are defined by~\eqref{eqDefEtazIedge}\&\eqref{eqDefEtamIedge}. Then, $u_a, u_b$ are solutions of the differential equation
\begin{equation} \label{Dirac_equation_formal}
  \big( \iu \partial \otimes \sigma_2 + m \otimes \sigma_3 \big) u = z u
\end{equation}
such that $u_a$ satisfies the boundary conditions in $\Dom \widetilde{D}^0$ at $a$ and $u_b$ satisfies the boundary conditions in $\Dom \widetilde{D}^0$ at $b$. Hence, by \cite[Satz~15.13]{WeidmannII} 
\begin{equation*}
  (\widetilde{D}^0 - z)^{-1} f(x) = \frac{1}{W(u_a,u_b)} \left(  u_b(x) \int_a^x \langle \overline{u_a(y)}, f(y) \rangle \dd y + u_a(x) \int_x^b \langle \overline{u_b(y)}, f(y) \rangle \dd y \right)
\end{equation*}
holds, where, with an arbitrary $x \in (a,b)$,
\begin{equation*}
  W(u_a,u_b) = \det (u_a(x), u_b(x)) = \begin{cases} (\alpha(z)\sin(k(z) (b-a)))^{-1}, & z \neq m, \\ ( 2 m (b-a))^{-1}, & z=m, \end{cases}
\end{equation*}
is the Wronskian for $u_a, u_b$. Assume now that $z\in\C\setminus((-\infty, -m] \cup [m, +\infty))$ (for the remaining $z$'s away from $\sigma(D^0)$ one can proceed similarly).  Then by our choice $\arg k(z) \in [0, \pi)$ we have $\overline{k(z)} = -k(\overline{z})$ and $\overline{\alpha(z)} = -\alpha(\overline{z})$, which implies for any $c \in \mathbb{R}$ (via their representations as power series)
\begin{equation*}
  \overline{e^{\iu k(z) c}} = e^{\iu k(\overline{z}) c}, \quad \overline{\sin(k(z) c)} = -\sin(k(\overline{z}) c), \quad \overline{\cos(k(z) c)} = \cos(k(\overline{z}) c).
\end{equation*}
This yields
\begin{equation*}
  U \overline{u_a(y)} = U \overline{U^* \eta_{z,2}(y)} = U U \overline{\eta_{z,2}(y)} = \eta_{\overline{z},2}(y) 
\end{equation*}
and likewise $U \overline{u_b(y)} = \eta_{\overline{z},1}(y)$. Therefore, we conclude
\begin{equation*}
  \begin{split}
    (D^0 &- z)^{-1} f(x) = U (\widetilde{D}^0 - z)^{-1} U^* f(x) \\
    &= \frac{1}{W(u_a,u_b)} \left(  U u_b(x) \int_a^x \langle U \overline{u_a(y)}, f(y)\rangle \dd y + U u_a(x) \int_x^b \langle U \overline{u_b(y)}, f(y) \rangle \dd y \right) \\
    &= \int_a^b R_z(x,y) f(y) \dd y.
  \end{split}
\end{equation*}
\end{proof}

Next, we compute the Krein $\gamma$-field and the Krein $\Kreinfun$-function induced by the boundary triple $( \bma, \bmb, \C^2 )$. Recall that for $z \in \rho(D^0)$ the functions $\eta_{z,1}, \eta_{z,2}$ are the basis of $\Ker(D^0-z)$ defined in~\eqref{eqDefEtazIedge}\&\eqref{eqDefEtamIedge}.

\begin{proposition} \label{proposition_gamma_internal_edge}
Let $D^{max}$ be defined by~\eqref{eqDefDiracMaxIedge}, $( \bma, \bmb, \C^2 )$ be the boundary triple  for $D^{max}$ given by~\eqref{eqDefBTIedge}, and $\refoperator = D^{max} \upharpoonright \Ker \Gamma^1$.
Then, the following is true:
\begin{itemize}
  \item[\textup{(i)}] The values of the induced Krein $\gamma$-field on $z \in \resolvent(\refoperator)$ are given by
\begin{equation*}
\gamma(z): \C^2 \rightarrow L^2((a,b);\C^2), \qquad 
\gamma(z) \generalVector{\omega}{}
=
\omega^1 \eta_{z,1} + \omega^2 \eta_{z,2}.
\end{equation*}
\item[\textup{(ii)}] The  adjoint of the value of the induced Krein $\gamma$-field at $z \in \rho(D^0)$ is given by
\begin{equation*}
\gamma^*(z): L^2 ( \iedge;\C^2 ) \to \C^2, \qquad 
\gamma^*(z) \phi
=
\begin{pmatrix}
\sclrp{\eta_{z,1}}{\phi} \\ \sclrp{\eta_{z,2}}{\phi}
\end{pmatrix}, \quad  (\forall \phi \in L^2 ( \iedge; \C^2 )).
\end{equation*}
\item[\textup{(iii)}] The values of the induced Krein $\Kreinfun$-function $\Kreinfun(z): \mathbb{C}^2\rightarrow \mathbb{C}^2$ at $z \in \rho(D^0) \setminus \{ m \}$ are given by
\begin{equation} \label{Weyl_function_internal_edge_z}
\Kreinfun(z)
=
\frac{-1}{\idk(z) \sink{(b-a)}}
\begin{pmatrix}
\cosk{(b-a)} & -1 \\
-1 & \cosk{(b-a)}
\end{pmatrix}
\end{equation}
and for $z=m$ by
\begin{equation} \label{Weyl_function_internal_edge_m}
\Kreinfun (m)
=
\frac{-1}{2m(b-a)}
\begin{pmatrix}
1 & -1 \\
-1 & 1
\end{pmatrix}.
\end{equation}
\end{itemize}
\end{proposition}
\begin{proof}
Note that $\eta_{z,1}$, $\eta_{z,2}$ defined by~\eqref{eqDefEtazIedge}\&\eqref{eqDefEtamIedge}  have the property that
\begin{equation*}
  \Gamma^1 \eta_{z,1} = \begin{pmatrix} 1 \\ 0 \end{pmatrix} \quad \text{and} \quad 
  \Gamma^1 \eta_{z,2} = \begin{pmatrix} 0 \\ 1 \end{pmatrix}.
\end{equation*}
Taking the definition \eqref{eq:gammaField} of the induced Krein $\gamma$-field into account, we find that this implies the claim of~(i).
  The claim in~(ii) is an immediate consequence of~(i).
 Finally, the claim in~(iii) follows directly from~(i) by applying $\Gamma^2$ to the expression for $\gamma(z)$.
\end{proof}

\subsection{Dirac operator on an external edge}\label{secEedge}

In this section we consider the situation that $I=(a,b)\subset\R$ with either $a=-\infty$ or $b=+\infty$. The former case will be labelled with $\orient=+1$ and the latter with $\orient=-1$. Later, the number $\orient$ will indicate whether the half-line, which models an unbounded edge in a graph, is entering or leaving a vertex. We denote the finite value of $a,b$ by $\gb$. We will employ a similar procedure as in Section~\ref{secIedge} to construct a boundary triple for Dirac operators in  $\HH = L^2(\eedge;\C^2)$. 

In a similar way as in Section~\ref{secIedge} we define for $m\geq 0$ the minimal realization $D^{min}$ and the maximal realization $D^{max}$ of the Dirac operator in $\HH$ by
\begin{equation}
\begin{aligned}\label{eqDefDiracMaxEedge}
D^{max} = \diracop, \qquad  \Dom D^{max} = H^{1}_{} ( \eedge; \C^2  ),
\end{aligned}
\end{equation}
and
\begin{equation*}
\begin{aligned}
D^{min} = \diracop, \qquad \Dom D^{min} = H^{1}_{0} ( \eedge; \C^2  ),
\end{aligned}
\end{equation*}
respectively.
Note that the minimal Dirac operator is again closed and symmetric, and the maximal Dirac operator is its adjoint 
$D^{max} = (D^{min})^*$. In the next proposition we introduce a boundary triple for $D^{max}$. Note that the same boundary triple for the Dirac operator on $(a,+\infty)$ was considered in \cite{CaMaPo_2013}. Therein, also the case of the half-line $(-\infty,b)$ is considered but with the roles of $\bma$ and $\bmb$ interchanged. We decided to deal with the both cases uniformly. For completeness, we provide all proofs.

\begin{proposition}
Define the two linear maps $\bma, \bmb: H^1 (\eedge; \C^2 ) \to \C$ by 
\begin{align}\label{eqDefBTEedge}
\bma \phi &= \phi^1(\gb), & \bmb \phi &= -\iu \orient \phi^2(\gb).
\end{align}
Then the triple $( \bma, \bmb, \C )$ is a boundary triple for $D^{max}$.
\end{proposition}
\begin{proof}
First, integration by parts yields the abstract Green identity \eqref{eq:Green}.

Next, to show the surjectivity of $(\bma, \bmb)$, consider the function $f(x) = \ee^{\orient \left( x - \gb \right)}$. Then, it is not difficult to see that 
\begin{equation*}
  \vectorComp{f \\ 0}, \vectorComp{0 \\ \iu \orient f} \in H^1(\eedge; \mathbb{C}^2)
\end{equation*}
are mapped by $( \bma, \bmb )$ to the standard basis of $\C^2$. This yields the surjectivity of $( \bma, \bmb )$. 

Finally, $D^{min}$ is densely defined, closed and symmetric with adjoint $(D^{min})^* = D^{max}$. Therefore, we conclude from Definition \ref{def:BT} that $( \bma, \bmb, \C )$ is a boundary triple for $D^{max}$.
\end{proof}

In the next proposition we compute the defect subspaces for $D^{max}$. Recall that the numbers $k(z)$ and $\alpha(z)$ are defined by~\eqref{def_k_z} and~\eqref{def_alpha}, respectively. 

\begin{proposition} \label{proposition_defect_space_external}
Let $D^{max}$ be defined by~\eqref{eqDefDiracMaxEedge}. Then, the following is true:
\begin{itemize}
  \item[\textup{(i)}] If $\Im \, k(z) > 0$, then $N_z = \Ker(D^{max} - z) = \textup{span}\, \{ \eta_z \}$, where
  \begin{equation}\label{eqDefEtazEedge}
\eta_{z}(x)
=
\ee^{-\iu \orient k(z) (x-\gb)} 
\vectorComp{ 1 \\ \frac{-\orient}{\idk(z)}}.
\end{equation}
\item[\textup{(ii)}] If $\Im \, k(z) = 0$, then $N_z = \Ker(D^{max} - z) = \{ 0 \}$.
\end{itemize}
\end{proposition}
\begin{proof}
We show the claim for $\eedge = (a, +\infty)$; the case $\eedge = (-\infty, b)$ can be treated similarly. Note that in this case $\rho=-1$. 

Let $z \in \mathbb{C}$ be fixed. As in~\eqref{eqsolutionOfDefectSubspaces} one finds that $N_z = \Ker(D^{max} - z)$ consists of all functions of the form
\begin{equation*}
\phi(x) = \exp\left[ \left( \iu z \pauli{1} - m \pauli{2} \right) x \right] {\omega}{}, \quad x \in (a,+\infty),
\end{equation*}
with $\omega \in \mathbb{C}^2$, which belong to $\Dom D^{max} = H^1((a,+\infty); \C^2)$. If  $k(z)\neq 0$, this expression can be rewritten as
\begin{equation}\label{eqNzDifferentialEquationSolutionEedge}
\phi(x) = 
\left[
\ee^{\iu k(z) x} \frac{1}{2} \left( I + A \right)
+
\ee^{-\iu k(z) x} \frac{1}{2} \left( I - A \right)
\right]
{\omega}{},
\end{equation}
where $A=\frac{z}{k(z)} \pauli{1} + \frac{\iu m}{k(z)} \pauli{2} $. Note that $A$ has the eigenvalues $\pm 1$. If $\Im \, k(z) >0$, the term with $\ee^{-i k(z) x}$ is not square integrable. Consequently ${\omega}{}$ has to be an eigenvector of $A$ for the eigenvalue $1$,
\begin{equation*}
\omega = \widetilde{\omega} e^{-\iu k(z) \gb} \vectorComp{ 1 \\ \frac{1}{\idk(z)} }, \quad \widetilde{\omega} \in \mathbb{C},
\end{equation*}
and  $\phi = \widetilde{\omega} \eta_z$ with $\eta_z$ given by~\eqref{eqDefEtazEedge}.  This completes the proof of (i).

It remains to prove item~(ii). Let $z\in\C$ be such that $\Im\, k(z)= 0$. If  $k(z)\neq 0$ then both exponential terms in~\eqref{eqNzDifferentialEquationSolutionEedge} are not integrable; and if $k(z) = 0 $ then the form of the solution is polynomial in $x$ and thus also not square integrable. Therefore, for $\Im\, k(z)= 0$ one has $N_z = \{ 0 \}$. This finishes the proof of the proposition.
\end{proof}

In the next proposition we compute the spectrum and the resolvent of $D^0 = D^{max} \upharpoonright {\Ker \bma}$, which is now given by
\begin{equation*}
D^0=\diracop, \quad  \Dom D^0 =\left\{\phi\in H^{1} ( \eedge; \C^2  )\vert \, \phi^1(\gb)=0\right\}.
\end{equation*}
Recall that the operator $D^0$ is self-adjoint in $\HH$, cf. Section~\ref{sec:BT}, and that its non-relativistic limit is either the Dirichlet or Neumann Laplacian, depending on whether we subtract the rest energy of a particle or an antiparticle during the limit procedure\cite{Cu_2014}. Also recall that $\eta_z$, $z \in \C \setminus ((-\infty, -m]\cup[m,+\infty))$, is defined by~\eqref{eqDefEtazEedge}. Moreover, we introduce for $z \in \C \setminus ((-\infty, -m]\cup[m,+\infty))$ the function
\begin{equation*}
  \mu_z(x) = \vectorComp{\sink{\rho (\partial-x)} \\ \frac{\rho \iu}{\idk(z)}\cosk{\rho(\partial-x)}}, \qquad x \in I.
\end{equation*}
Note that the evaluation of $\mu_z$ coincides for $I = (a, +\infty)$ with $\sink{(b-a)} \eta_{z,2}$ and for $I = (-\infty, b)$ with $\sink{(b-a)} \eta_{z,1}$ defined by~\eqref{eqDefEtazIedge}; however, since $\eta_{z,1}$ and $\eta_{z,2}$ are elements of $L^2(\iedge; \C^2)$, we use a different symbol here.

\begin{proposition} \label{proposition_D_0_external_edge}
Let $\refoperator = D^{max} \upharpoonright {\Ker \bma} $. Then, the following is true:
\begin{itemize}
  \item[\textup{(i)}] $\sigma(D^0)$ is purely continuous and given by
\begin{equation*}
\spectrum (\refoperator)
=
\spectrum_{c} (\refoperator)
=
\left( - \infty, -m \right] \cup \left[ +m, + \infty \right).
\end{equation*}

\item[\textup{(ii)}] Define for $z \in \rho(D^0)$ the function $R_z: \eedge \times \eedge \rightarrow \C^{2 \times 2}$ by
\begin{align*}
R_z(x,y)
&=
\begin{cases}
	-\alpha(z) \eta_z(x) \mu_{\overline{z}}(y)^*, & \orient \left(x-y\right) < 0, \\
	\alpha(z) \mu_z(x) \eta_{\overline{z}}(y)^*, & \orient \left(x-y\right) > 0.
\end{cases} 
\end{align*}
Then, the resolvent of $D^0$ acts as
\begin{equation*}
(D^0 - z)^{-1} \phi(x)
=
\int_{\eedge} R_z(x,y) \phi(y) \dd y, \qquad \phi \in L^2(\eedge; \C^2).
\end{equation*}
\end{itemize}
\end{proposition}
For $I = (a, +\infty)$ the results in Proposition~\ref{proposition_D_0_external_edge} are also stated without proof in \cite[Lemma~3.3]{CaMaPo_2013}; for completeness, we provide a simple proof here.
\begin{proof}[Proof of Proposition~\ref{proposition_D_0_external_edge}]
  For the proof of (i), we will make use of the relation
  \begin{equation*} 
    (D^0)^2 = \begin{pmatrix} m^2 -\Delta_D & 0 \\ 0 & m^2 - \Delta_N \end{pmatrix},
  \end{equation*}
  where $-\Delta_D$ and $-\Delta_N$ denote the realizations of the Laplace operator on $\eedge$ with Dirichlet and Neumann boundary conditions, respectively, which follows from a direct calculation. Therefore, as $-\Delta_D$ and $-\Delta_N$ are both non-negative, the spectral mapping theorem implies that
  \begin{equation} \label{inclusion_spectrum_D_0_external1}
    \spectrum (\refoperator) \subset \left( - \infty, -m \right] \cup \left[ +m, + \infty \right).
  \end{equation}
  In particular, for all $z \in \sigma(D^0)$ one has for the number $k(z)$ defined by~\eqref{def_k_z} that $\Im \, k(z) = 0$ and therefore, it follows from Proposition~\ref{proposition_defect_space_external} that $\sigma_p(D^0) = \emptyset$ and hence, as $D^0$ is self-adjoint, $\sigma(D^0) = \sigma_c(D^0)$.

  To show the inverse inclusion in~\eqref{inclusion_spectrum_D_0_external1}, consider for $z \in (-\infty, -m] \cup [+m, +\infty)$ and $n \in \mathbb{N}$ the functions
  \begin{equation*}
    \phi_n(x) := \frac{1}{\sqrt{n}} \chi\left( \frac{1}{n} |x - \gb + \rho n^2| \right) \ \ee^{\iu k(z) x} \big( k(z) \sigma_1 + m \sigma_3 + z \sigma_0 \big) \omega,
  \end{equation*}
  where $\chi \in C^\infty(\mathbb{R})$ such that $\chi(r) = 1$, if $|r| < \frac{1}{2}$, and $\chi(r) = 0$, if $|r| > 1$, and $\omega \in \mathbb{C}^2$ such that $( k(z) \sigma_1 + m \sigma_3 + z \sigma_0 ) \omega \neq 0$. Then, one verifies that, for all sufficiently large $n$,  $\phi_n \in \Dom D^0$, $\| \phi_n \| = $ const., and $\|(D^0 - z) \phi_n\| \rightarrow 0$, as $n \rightarrow +\infty$. Therefore, $z \in \sigma(D^0)$ and the reverse inclusion in~\eqref{inclusion_spectrum_D_0_external1} is also shown.
  
  The proof of~(ii) follows the same lines as the one of Proposition~\ref{proposition_D_0_internal_edge}~(ii) using \cite[Satz~15.17]{WeidmannII} instead of \cite[Satz~15.13]{WeidmannII} with, for $I = (a, +\infty)$, the functions $u_a = U^* \mu_z$ and $u_b = U^* \eta_z$, which are solutions of the formal Dirac equation~\eqref{Dirac_equation_formal} that satisfy the boundary condition at $a$ and $b=+\infty$, respectively; for $I = (-\infty, b)$ one can use $u_a = U^* \eta_z$ and $u_b = U^* \mu_z$. The details are left to the reader.
\end{proof}

Now, we are prepared to compute the Krein $\gamma$-field and the Krein $\Kreinfun$-function induced by the boundary triple $( \bma, \bmb, \C )$. For $I = (a, +\infty)$ this result is also stated without proof in \cite[Lemma~3.3]{CaMaPo_2013}; again, we provide for completeness a short and simple proof.
Recall that $\eta_z$, $z \in \rho(D^0)$, is the function defined by~\eqref{eqDefEtazEedge}.

\begin{proposition}  \label{proposition_gamma_external_edge}
Let $D^{max}$ be defined by~\eqref{eqDefDiracMaxEedge}, $( \bma, \bmb, \C^2 )$ be the boundary triple  for $D^{max}$ given by~\eqref{eqDefBTEedge}, and $\refoperator = D^{max} \upharpoonright \Ker \Gamma^1$.
Then, the following is true:
\begin{itemize}
  \item[\textup{(i)}] The values of the induced Krein $\gamma$-field on $z \in \resolvent(\refoperator)$ are given by
\begin{equation*}
\gamma(z): \C \rightarrow L^2(\eedge; \C^2), \qquad 
\gamma(z) \omega
=
\omega \eta_{z}.
\end{equation*}
\item[\textup{(ii)}] The  adjoint of the values of the induced Krein $\gamma$-field at $z \in \rho(D^0)$ is given by
\begin{equation*}
\gamma^*(z): L^2 ( \eedge; \C^2  ) \rightarrow \C, \qquad
\gamma^*(z) \phi =
 \sclrp{\eta_z}{\phi} .
\end{equation*}
\item[\textup{(iii)}] The values of the induced Krein $\Kreinfun$-function  $\Kreinfun(z): \mathbb{C}\rightarrow \mathbb{C}$ at $z \in \rho(D^0)$ are given by
\begin{equation}\label{eqKreinQEedge}
\Kreinfun(z) = 
\frac{\iu}{\idk(z)}.
\end{equation}
\end{itemize}
\end{proposition}
Note that the induced Krein $\Kreinfun$-function in~\eqref{eqKreinQEedge} does not depend on the orientation of $\eedge$.
\begin{proof}[Proof of Proposition~\ref{proposition_gamma_external_edge}]
The claim in~(i) follows immediately from the definition \eqref{eq:gammaField} of $\gamma(z)$ and the fact that $\eta_z \in N_z$ satisfies $\Gamma^1 \eta_z = 1$. The claim in~(ii) is an immediate consequence of~(i).
  Finally, the claim in~(iii) follows directly from~(i) by applying $\Gamma^2$ to the expression for $\gamma(z)$.
\end{proof}

\subsection{Dirac operator on a graph}\label{sec:graph}
Let $\graph \equiv \left( \vertices, \edges, \gb \right)$ be an oriented graph endowed with the parametrization as described in Section \ref{sec:graphs}. Recall that the spaces $L^2(\graph; \C^2)$, $\widetilde{H}^1(\graph; \C^2)$, and $\widetilde{H}^1_0(\graph; \C^2)$ are defined by~\eqref{def_L_2_graph} and~\eqref{def_Sobolev_space_graph}, respectively.
We start this section by introducing a minimal and a maximal realization of the Dirac operator in it. As described in Section~\ref{sec:graphs} every edge $j\in\edges$ is associated with an interval $I_j$ and with a corresponding Hilbert space $\HH_{j} = L^2(I_j ; \C^2 )$.
On every edge $j$ we introduce the minimal Dirac operator $D_j^{min}$ and the maximal Dirac operator $D_j^{max}$ as in Sections \ref{secIedge} and \ref{secEedge}.
The minimal and the maximal Dirac operator on the whole graph are defined as direct sums of the corresponding operators on the edges with respect to the direct sum decomposition of $\HH := L^2(\graph; \C^2)$,
\begin{equation*}
D^{min} 
:= \bigoplus_{j\in \edges} D_j^{min}, 
\quad  \quad
\Dom D^{min} := \bigoplus_{j\in\edges} H^1_0 \left( I_j ; \C^2 \right) \equiv \widetilde{H}^1_0 \left(\graph; \C^2 \right),
\end{equation*}
and
\begin{equation*}
D^{max} 
:= \bigoplus_{j\in \edges} D_j^{max},
\quad  \quad
\Dom D^{max} := \bigoplus_{j\in\edges} H^1 \left( I_j ; \C^2 \right) \equiv \widetilde{H}^1 \left(\graph; \C^2 \right).
\end{equation*}
Since we are dealing with finite direct sums, it follows from the properties of $D^{min}_j$ and $D^{max}_j$, $j \in \edges$, that $D^{min}$ is a closed and symmetric operator in $\HH$ and that $(D^{min})^* = D^{max}$.

We will also make use of the reference operator $\refoperator = \bigoplus_{j\in\edges} \refoperator_{j}$, which is given explicitly by
\begin{equation} \label{def_D_0_graph}
\begin{split}
  (D^0 \Phi)_j &= (\diracop) \phi_j \qquad \forall j \in \edges, \\
  \Dom D^0 &= \Big\{ \Phi = (\phi_j)_{j \in \edges} \in \widetilde{H}^1(\graph; \C^2) \, | \,\phi^{1}_e(\partial_{\rho(e)} e) = 0 \,\, \forall e \in \extes,  \\
  &\qquad \qquad \qquad \qquad \qquad \qquad \qquad  \phi_i^1(\partial_+ i) = \phi_i^1(\partial_- i) = 0 \,\, \forall i  \in \intes \Big\}.
\end{split}
\end{equation}
It follows directly from the definition that
\begin{equation*}
\spectrum(\refoperator) = \bigcup_{j\in\edges} \spectrum (\refoperator_j)
\end{equation*}
and the spectrum of each $D^0_j$ is described in Propositions~\ref{proposition_D_0_internal_edge} and~\ref{proposition_D_0_external_edge}, i.e., we get the following result.

\begin{proposition} \label{prop:spec_D^0}
For any finite graph, $\sigma(D^0)\subset(-\infty,-m]\cup[m,+\infty)$. The equality holds if and only if $\extes\neq \emptyset$ in which case $\sigma(D^0)=\sigma_{ess}(D^0)$. If $\extes=\emptyset$ then $\sigma(D^0)$ is purely discrete.
\end{proposition}

Likewise, the resolvent of $D^0$ has a direct sum structure,
\begin{equation*}
  (D^0 - z)^{-1} = \bigoplus_{j \in \edges} (D^0_j - z)^{-1} \qquad \big(\forall z \in \rho(D^0) = \bigcap_{j\in\edges} \rho (\refoperator_j)\big),
\end{equation*}
and the resolvents of $D^0_j$ are the integral operators described in the previously mentioned propositions.
Let us now put
\begin{equation} \label{eq:Gdecomp}
\boundspace := \bigoplus_{j\in\edges} \boundspace_j,
\quad\quad 
\boundspace_j:=
\begin{cases}
\C^2, & j\in\intes, \\
\C, & j\in\extes,
\end{cases}
\end{equation}
and, for every edge $j\in\edges$, label the boundary mappings introduced in \eqref{eqDefBTIedge} and \eqref{eqDefBTEedge} with a subscript $j$. Then $(\bma_j,\bmb_j,\boundspace_j)$ is a boundary triple for $D^{max}_j$. Taking the convention on the evaluation of  $\Phi = (\phi_j)_{j \in \edges} \in \widetilde{H}^1(\graph; \C^2)$ at $v \in \vertices$ from the end of Section~\ref{sec:graphs} into account, we can write for $j=i\in\intes$
\begin{equation*}
\bma_i \phi_i = \vectorComp{ \phi_i^1(\gb_- i) \\ \phi_i^1 (\gb_+ i) }, \qquad
\bmb_i \phi_i = \vectorComp{\iu \phi_i^2 (\gb_-i ) \\ - \iu \phi_i^2 (\gb_+ i)}, 
\end{equation*}
and for $j=e\in\extes$
\begin{equation*}
\bma_e \phi_e = \vectorComp{ \phi_e^1 (\gb_{\orient(e)} e )}, \qquad
\bmb_e \phi_e = \vectorComp{ -  \iu \orient(e) \phi_e^2 (\gb_{\orient(e)} e) }. 
\end{equation*}
In a similar way as above, we denote the Krein $\gamma$-field and the Krein $\Kreinfun$-function induced by the boundary triple $(\bma_j,\bmb_j,\boundspace_j)$ by $\gamma_j$ and $\Kreinfun_j$, respectively; cf. Propositions~\ref{proposition_gamma_internal_edge} and~\ref{proposition_gamma_external_edge}. Due to the direct sum structure of all the objects involved, we get immediately from Theorem~\ref{theorem_direct_sum} the following result.

\begin{proposition} \label{prop:totalBT}
Let $\bma, \bmb: \widetilde{H}^1(\graph;\C^2) \to \boundspace$ be defined by
\begin{equation*}
\bma := \bigoplus_{j\in\edges} \bma_j,
\quad\quad
\bmb := \bigoplus_{j\in\edges} \bmb_j.
\end{equation*}
Then  $( \bma, \bmb, \boundspace )$ is a boundary triple for $D^{max}$ such that $D^{max} \upharpoonright \Ker \bma = D^0$ with $D^0$ given by~\eqref{def_D_0_graph}. Moreover, the induced Krein $\gamma$-field is given by
\begin{equation*}
\gamma(z) = \bigoplus_{j\in\edges} \gamma_j(z): \boundspace \rightarrow \HH, \qquad z \in \rho(D^0),
\end{equation*}
and the induced Krein $\Kreinfun$-function is
\begin{equation*}
\Kreinfun(z)=\bigoplus_{j\in\edges} \Kreinfun_j (z): \boundspace \rightarrow \boundspace, \quad z \in \rho(D^0).
\end{equation*}

\end{proposition}

In the rest of this section we construct various realizations of the Dirac operator on the graph, employing the framework from Section \ref{sec:BT}, i.e., by restricting $D^{max}$ to
\begin{equation} \label{eq:DLambdaRest}
 \{\phi\in \widetilde{H}^1(\graph,\C^2)\vert\,(\bma\phi,\bmb\phi)\in\Lambda\},
\end{equation}
where $\Lambda$ is a relation in $\boundspace$. We will be interested in physically more relevant \emph{local} transmission conditions that do not mix boundary values at different vertices. To describe these conditions, recall that for $v \in \vertices$ the vertex space $\vertspace_v$ and $\vertspace$ are defined in~\eqref{eq:Fdecomp}.
Note that $\vertspace$ is isomorphic to $\C^{\sum_{v\in\vertices} \deg v}$ and the boundary space is isomorphic to $\C^{2 \left\vert \intes \right\vert + \left\vert \extes \right\vert}$. Employing the handshake lemma, which says that $\sum_{v\in\vertices} \deg v  =  2 \left\vert \intes \right\vert + \left\vert \extes \right\vert$, we conclude $\vertspace \simeq \boundspace$, i.e., we can identify these spaces. 

The local transmission conditions will be described by relations $\Lambda$ in $\vertspace$ that decompose with respect to the decomposition \eqref{eq:Fdecomp} of the vertex space $\vertspace$, i.e., they are direct sums of relations $\Lambda_v$ in $\vertspace_v$. 
On the other hand, the boundary mappings in \eqref{eq:DLambdaRest} decompose to the direct sum with respect to the decomposition \eqref{eq:Gdecomp} of the boundary space $\boundspace$. To overcome this discrepancy, we will construct a linear isomorphism $W:\boundspace\to\vertspace$ that takes appropriately one arrangement of $\C^N$, $N:=2 \left\vert \intes \right\vert + \left\vert \extes \right\vert$, to another. The idea is that both spaces collect complex numbers at all evaluation points of a function on the graph. The boundary space does it edge by edge and the vertex space does it vertex by vertex. 

In order to write $W$ explicitly, it is necessary to agree on two numberings of the evaluation points, $\nu_\boundspace$ for the boundary space, $\nu_\vertspace$ for the vertex space. By \emph{numbering} it is meant here any injective map $\nu: \ito{N} \to \edges \times \vertices$ with the image $\set{(j,v)}{v\in\gb j}$. The sought linear isomorphism $W:\boundspace \to \vertspace$ is then given by a unitary matrix $W\in\C^{N \times N}$ with elements
\begin{equation} \label{def_W}
W_{n,m}=\begin{cases}
1, & \nu_\vertspace (n) = \nu_\boundspace (m), \\
0, & \text{otherwise},
\end{cases} \qquad (\forall n,m \in \ito{N}).
\end{equation}
In other words, it is a permutation matrix for the permutation $\nu_\vertspace^{-1}\nu_\boundspace$.
The convention, adopted in this article, is that the total orders on the set of edges and the set of vertices, used in~\eqref{def_incidence_matrix} in the definition of the incidence matrix $G$, are reused for the numberings. The \emph{edge-vertex} numbering $\nu_\boundspace$ preserves the order of edges, for $e\in\extes$ there is then a unique element $(e,\gb_{\orient(e)}e)$, for $i\in\intes$ the preimage of the couple with the initial vertex precedes the couple with the terminal vertex $\nu_\boundspace^{-1}\left(\left(i,\gb_- i\right)\right)+1=\nu_\boundspace^{-1}\left(\left(i,\gb_+ i\right)\right)$. On the other hand, the \emph{vertex-edge} numbering $\nu_\vertspace$ visits the vertices in the order of $\vertices$ and at each vertex $v$ it follows the order of edges restricted to $\edges_v$. The same result is achieved by constructing the colexicographic order on $\set{(j,v)}{v\in\gb j}$.

\begin{remark}
It is possible to describe an algorithm how to construct $W$ defined by~\eqref{def_W} from the incidence matrix $G$ given by~\eqref{def_incidence_matrix}. By definition, each column in $G$ contains exactly one or two non-zero elements. Columns with exactly one non-zero element correspond to unbounded edges. Replace every non-zero elements in them with 1, leave the zeros, i.e., use the substitution $\{-1 \to 1, 1 \to 1, 0\to 0\}$.
Columns with two non-zero elements are  replaced with two columns, if there was $-1$, place $1$ to the left column, if there was $1$, place $1$ to the right column. The other entries will be zeros. In other words, in columns with two non-zero entries apply the substitution $\{-1 \to (1\quad 0), 1 \to (0 \quad 1) , 0 \to (0 \quad 0) \}$. Next, each row has to be replaced by a number of rows equal to the number of non-zero elements in that row. The first replacing row consists of zeros except for the first position of non-zero element from the left where a $1$ will be. The second replacing row will carry over the second non-zero position from the left and so on.
For example, for a graph depicted in Figure \ref{fig:graph} we have

\begin{figure}[h] 
\centering
\includegraphics[width=200px]{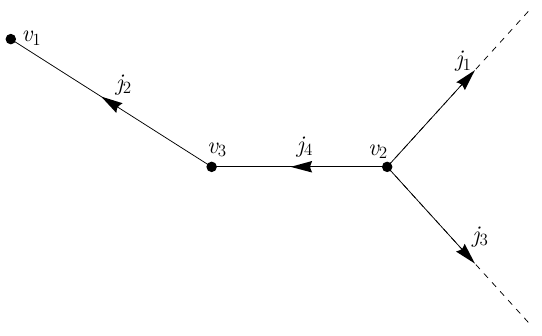} 
\caption{A graph with three vertices $(v_1,v_2,v_3)$ and four edges $(j_1,\ldots,j_4)$ whose incidence matrix $G$ is given in \eqref{eq:incidence_ex}.}
\label{fig:graph}
\end{figure}

\begin{align} \label{eq:incidence_ex}
G &= \begin{pmatrix}
0 & 1 & 0 & 0 \\
-1 & 0 & -1 & -1 \\
0 & -1 & 0 & 1
\end{pmatrix},
&
W &= \begin{pmatrix}
0 & 0 & 1 & 0 & 0 & 0 \\
1 & 0 & 0 & 0 & 0 & 0 \\
0 & 0 & 0 & 1 & 0 & 0 \\
0 & 0 & 0 & 0 & 1 & 0 \\
0 & 1 & 0 & 0 & 0 & 0 \\
0 & 0 & 0 & 0 & 0 & 1 \\
\end{pmatrix}.
\end{align}
\end{remark}

With the help of the isomorphism $W$ one can relate uniquely a given linear relation $\Lambda$ in $\vertspace$ to a relation in $\boundspace$ via 
\begin{equation} \label{def_Lambda_W}
  \Lambda_W := \{ (f, f') \in \boundspace \oplus \boundspace \, | \, (W f, W f') \in \Lambda \}.
\end{equation}
In particular, we associate with a pair $A,B$ of linear mappings in $\vertspace$ the following linear relation $\Lambda_{A,B}$ in $\boundspace$,
\begin{equation} \label{eq:def_Lambda_AB}
  \Lambda_{A,B} := \{ (f, f') \in \boundspace \oplus \boundspace \, | \, AW f=BW f'\}.
\end{equation}
To summarize, we arrived at the following definition.

\begin{definition} \label{def:Dirac_on_graph}
Let $\graph = \left( \vertices, \edges, \gb \right)$ be an oriented graph with the vertex space $\vertspace = \bigoplus_{v\in\vertices} \vertspace_v$.
A linear relation $\Lambda$ in $\vertspace$ is called local if there are linear relations $\Lambda_v$ in $\vertspace_v$ such that $\Lambda = \bigoplus_{v\in\vertices} \Lambda_v$. Given a local linear relation $\Lambda$ in $\vertspace$ and its equivalent $\Lambda_W$ in $\boundspace$ given by~\eqref{def_Lambda_W}, we define the corresponding Dirac operator $D^{\Lambda_W}$ with local transmission conditions as the restriction of $D^{max}$ to
\begin{equation}\label{DLambdaDomain} 
\Dom D^{\Lambda_W}
=
\set{\Phi \in \widetilde{H}^1(\graph;\C^2)}{ \left( W\bma\Phi, W\bmb\Phi \right) \in \Lambda }.
\end{equation}
If, in particular, $\Lambda_v = 
\set{ \left( f_1, f_2 \right) \in \vertspace \oplus \vertspace }{ A_v f_1 = B_v f_2}$
with a pair  $A_v, B_v$ of linear operators in $\vertspace_v$,
then $\Lambda$ is parametrized by the pair of operators $A = \bigoplus_{v\in\vertices} A_v$, $B = \bigoplus_{v\in\vertices} B_v$ in $\vertspace$ and its equivalent in $\boundspace$ is given by~\eqref{eq:def_Lambda_AB}. The associated Dirac operator $D^{\Lambda_{A,B}}$ is then the restriction of $D^{max}$ to
\begin{equation}\label{DLABDomain}
\Dom D^{\Lambda_{A,B}}
=
\set{\Phi \in \widetilde{H}^1(\graph;\C^2)}{ AW\bma\Phi = BW\bmb\Phi }. 
\end{equation}

\end{definition}

With the help of the boundary triple $(\bma, \bmb, \boundspace)$ we can describe all self-adjoint realizations $D^{\Lambda_{A,B}}$. Of course, for general (non-local) relations $\Lambda_{A,B}$ given by~\eqref{eq:def_Lambda_AB} Proposition~\ref{proposition_Lambda_self_adjoint} can be applied directly.
In the following proposition we discuss that the self-adjoint local boundary conditions can be characterized in a similar way also locally. Since the matrix $W$ is unitary, the following result is an immediate consequence of Proposition~\ref{proposition_Lambda_self_adjoint}, the fact that $D^{\Lambda_{A,B}}$ is self-adjoint if and only if $\Lambda_{A,B}$ is self-adjoint, and the direct sum structure of local relations in $\vertspace$.

\begin{proposition}
Let $\Lambda$ be a local relation in $\vertspace$ such that $\Lambda_W=\Lambda_{A,B}$  and let $D^{\Lambda_{A,B}}$ be defined by~\eqref{DLABDomain}. Then $D^{\Lambda_{A,B}}$ is self-adjoint in $\HH$ if and only if for all $v \in \vertices$ there holds $A_vB_v^*=B_vA_v^*$ and the block matrices  $(A_v\vert B_v)$ have maximal rank.
\end{proposition}

With the help of the boundary triple $(\bma, \bmb, \boundspace)$ from Proposition~\ref{prop:totalBT} and Theorem~\ref{theorem_Birman_Schwinger} one can now characterize spectral properties of the realizations $D^{\Lambda_{A,B}}$. We remark that the following result is true in exactly the same form for general (non-local) relations $\Lambda_{A,B}$ given by~\eqref{eq:def_Lambda_AB}. Recall that $D^0 = D^{max} \upharpoonright \Ker \bma$ is given by~\eqref{def_D_0_graph} and that $\gamma$ and $\Kreinfun$ are the Krein $\gamma$-field and the Krein $\Kreinfun$-function induced by the triple $(\bma, \bmb, \boundspace)$.

\begin{theorem} \label{theo:spectrum_AB}
Let $\Lambda$ be a local relation in $\vertspace$ such that $\Lambda_W=\Lambda_{A,B}$,  $D^{\Lambda_{A,B}}$ be given by~\eqref{DLABDomain}, and $W:\boundspace\to\vertspace$ be the linear isomorphism defined by~\eqref{def_W}.
Then for $z\in\resolvent(\refoperator)$ the following holds:
\begin{enumerate}
\item[\textup{(i)}] $z \in \spectrum_{p} \left( D^{\Lambda_{A,B}} \right)$ if and only if $0\in\spectrum \left( A -  B W \Kreinfun(z) W^{-1} \right)$.
\item[\textup{(ii)}] $z \in \resolvent \left( D^{\Lambda_{A,B}} \right)$ if and only if $0\in\resolvent \left(A -  B W \Kreinfun(z) W^{-1} \right)$. In this case,
\begin{equation} \label{eq:Krein_AB}
( D^{\Lambda_{A,B}} - z )^{-1}
= ( \refoperator - z )^{-1} +
\gamma(z) \left(A W -  BW \Kreinfun (z)  \right)^{-1}  B W \gamma(\cc{z})^*.
\end{equation}
\end{enumerate}
\end{theorem}

Using the resolvent formula \eqref{eq:Krein_AB}, we will immediately deduce the following stability result for the essential spectrum, which is again true for both local and non-local transmission conditions.
\begin{proposition}
If we have either
\begin{enumerate}
\item[\textup{(i)}] $(\extes=\emptyset \vee m>0)$ and $\rho(D^{\Lambda_{A,B}})\neq \emptyset$

\noindent or
\item[\textup{(ii)}] $\rho(D^{\Lambda_{A,B}})\cap\C^\pm\neq\emptyset$,
\end{enumerate}
then 
\begin{equation} \label{eq:ess_spec}
\sigma_{ess}(D^{\Lambda_{A,B}})=\sigma_{ess}(D^0)=
\begin{cases}
\emptyset, & \text{ if } \extes=\emptyset,\\
(-\infty,-m]\cup[m,+\infty), & \text{ if } \extes\neq\emptyset.
\end{cases}
\end{equation}
In particular, \eqref{eq:ess_spec} is always true when $D^{\Lambda_{A,B}}$ is self-adjoint.
\end{proposition}

\begin{proof}
Since, by  \eqref{eq:Krein_AB}, the difference $( D^{\Lambda_{A,B}} - z )^{-1}-( D^0 - z )^{-1}$ is a finite rank (and thus compact) operator, we may apply the Weyl essential spectrum theorem \cite[Theorem XIII.14]{RS4} which yields $\sigma_{ess}(D^{\Lambda_{A,B}})=\sigma_{ess}(D^0)$ under (ii) or $\sigma(D^0)\neq \R \wedge \rho(D^{\Lambda_{A,B}})\neq \emptyset$. By Proposition \ref{prop:spec_D^0},  $\sigma(D^0)\neq \R$ if and only if $\extes=\emptyset \vee m>0$. The formula for $\sigma_{ess}(D^0)$ follows from the same proposition. Finally, a self-adjoint $D^{\Lambda_{A,B}}$ always satisfies (ii).
\end{proof}

\subsection{Star shaped graph} \label{sec:star_shaped}
Let us demonstrate our results with an example of a star shaped graph that consists of a single vertex and $N\in\N$ outgoing external edges $(e_l)_{l=1}^N$, each of them parametrized by $(0,+\infty)$. Then $\boundspace=\C^N$, the boundary mappings may be chosen in the following way
\begin{equation*}
\bma \phi 
=
\vectorComp{
\phi^1_{e_1}(0) \\ \phi^1_{e_2}(0) \\ \vdots \\ \phi^1_{e_N}(0)
},
\quad
\bmb \phi 
=
\vectorComp{
\iu \phi^2_{e_1}(0) \\ \iu \phi^2_{e_2}(0) \\ \vdots \\ \iu \phi^2_{e_N}(0)
}
\qquad
\big(\forall \phi \in \widetilde{H}^1(\graph;\C^2)\big),
\end{equation*}
the induced Krein $\Kreinfun$-function reads
\begin{equation} \label{eq:starKrein}
\Kreinfun(z) = \frac{\iu}{\idk(z)} I_N,
\end{equation}
where $I_N$ stands for $N\times N$ identity matrix, and $W=I_N$. Finally, note that $\sigma(D^0)=(-\infty,-m]\cup[m,+\infty)$.

Applying Theorem \ref{theo:spectrum_AB}(i) we will deduce very explicit results for the spectrum of $D^{\Lambda_{A,B}}$ in terms of eigenvalues of the linear pencil  
$L(\lambda):=A-\lambda B$, $\lambda \in \C$,
defined by $A$ and $B$. Recall that, by definition, $\lambda$ belongs to the spectrum of $L$, that will be denoted by $\sigma(A,B)$, if and only if $0\in\sigma(L(\lambda))$. In particular, $\lambda$ is an eigenvalue of $L$ if $L(\lambda)$ is not invertible. In our finite dimensional setting, $\sigma(A,B)$ is exactly the set of eigenvalues of $L$ or, equivalently, 
$$\sigma(A,B)=\{\lambda\in\C|\, \det(L(\lambda))=\det(A-\lambda B)=0\}.$$

\begin{proposition} \label{prop:star_shaped}
Let $\graph$ be the star shaped graph described above, $A,B\in\C^{N\times N}$, and  $L(\lambda):=A-\lambda B$, $\lambda\in\C$, be the linear pencil defined by $A$ and $B$. Then  the following holds:
\begin{enumerate}
\item[\textup{(I)}] If $L$ is singular, i.e., $\det(L(\lambda))=0$ for all $\lambda\in\C$ or equivalently $\sigma(A,B)=\C$, then $\sigma(D^{\Lambda_{A,B}})=\C$ and $\C\setminus\sigma(D^0)\subset\sigma_p(D^{\Lambda_{A,B}})$.
\item[\textup{(II)}] If $L$ has no eigenvalues, i.e., $\det(L(\lambda))\neq 0$ for all $\lambda\in\C$ or equivalently $\sigma(A,B)=\emptyset$, then no $z\in\C\setminus\sigma(D^0)$ belongs to $\sigma(D^{\Lambda_{A,B}})$.
\item[\textup{(III)}] If $L$ is not singular but has some eigenvalues, i.e., $\det(L(\lambda))=0$ exactly for $\lambda\in\sigma(A,B)=\{\lambda_k\}_{k=1}^{\tilde{N}}$, where $1\leq\tilde{N}\leq N$, then one has:
\begin{enumerate}
\item[\textup{(i)}] If $\pm\iu\in\sigma(A,B)$ and $m=0$, then $\C^\pm\subset{\sigma_p(D^{\Lambda_{A,B}})}$ and $\overline{\C^\pm}\subset{\sigma(D^{\Lambda_{A,B}})}$; for $m>0$ there are no eigenvalues of $D^{\Lambda_{A,B}}$ corresponding to $\lambda_k=\pm\iu$.
\item[\textup{(ii)}] If $\Re(\lambda_k)<0$ and $m>0$ then
$$m\frac{1-\lambda_k^2}{1+\lambda_k^2}\in\sigma_p(D^{\Lambda_{A,B}});$$
in all other cases there are no eigenvalues of $D^{\Lambda_{A,B}}$ corresponding to $\lambda_k\neq\pm\iu$.
\end{enumerate}
\end{enumerate}
\end{proposition}

\begin{proof}
Take any $z\in\C\setminus\sigma(D^0)=\C\setminus((-\infty,-m]\cup[m,+\infty))$. Using \eqref{eq:starKrein} and  $W=I_N$,  the Birman Schwinger principle from Theorem \ref{theo:spectrum_AB}(i) yields that $z\in\sigma_p(D^{\Lambda_{A,B}})$ if and only if
$$0\in\sigma(A-BW\Kreinfun(z)W^{-1})=\sigma(A-\iu \alpha(z)^{-1}B),$$
which occurs if and only if 
\begin{equation*} 
\det(A-\iu\alpha(z)^{-1}B)=\det(L(\iu\alpha(z)^{-1}))=0
\end{equation*}
or equivalently $\iu\alpha(z)^{-1}\in\sigma(A,B)$.
This together with the closedness of the spectrum shows immediately \textup{(I)} and \textup{(II)}. 

To prove \textup{(III)}, we start with $\lambda\in\sigma(A,B)$ and look for $z\in\C\setminus\sigma(D^0)$ that obeys
\begin{equation} \label{star_spec_cond}
\lambda=\frac{\iu}{\alpha(z)}=\iu\frac{\sqrt{z^2-m^2}}{z+m}.
\end{equation}
Squaring this equation, we get the following necessary condition on $z$ to be in $\sigma_p(D^{\Lambda_{A,B}})$,
\begin{equation} \label{star_spec_cond2}
 (1+\lambda^2)z=m(1-\lambda^2).
\end{equation}
For $\lambda=\pm\iu$, \eqref{star_spec_cond2} can be only satisfied if $m=0$, in which case any $z\in\C$ is a solution. Plugging $m=0$ and $\lambda=\pm\iu$ into \eqref{star_spec_cond}, we get $\pm 1=\sgn(\Im(z))$. We conclude that if $m=0$ and $\iu$ or $-\iu$ belongs to $\sigma(A,B)$ then $\C^+$ or $\C^-$, respectively, belongs to $\sigma_p(D^{\Lambda_{A,B}})$. This together with the closedness of the spectrum yields \textup{(i)}. 

Finally, let $\lambda\in\sigma(A,B)\setminus\{-\iu,\iu\}$. Then we may divide in \eqref{star_spec_cond2} to get
\begin{equation} \label{eq:star_z}
 z=m\frac{1-\lambda^2}{1+\lambda^2}.
\end{equation}
It remains to check whether this solution satisfies also the original equation \eqref{star_spec_cond} and obeys $z\in\C\setminus\sigma(D^0)$.
This can not happen when $m=0$, so assume from now on that $m\neq 0$. Substituting \eqref{eq:star_z} into \eqref{star_spec_cond}, we arrive at
\begin{equation*}
\frac{\lambda}{1+\lambda^2}=\iu\sqrt{-\Big(\frac{\lambda}{1+\lambda^2}\Big)^2},
\end{equation*} 
which is only valid if
\begin{equation*}
\Re\Big(\frac{\lambda}{1+\lambda^2}\Big)<0 \qquad \text{or} \qquad \Re\Big(\frac{\lambda}{1+\lambda^2}\Big)=0 \,\wedge\,\Im\Big(\frac{\lambda}{1+\lambda^2}\Big)\geq 0.
\end{equation*}
In terms of $\lambda\neq\pm\iu$, this can be written as
\begin{equation*}
\Re(\lambda)<0 \qquad \text{or} \qquad \Re(\lambda)=0\,\wedge\,\Im(\lambda)\in(-\infty,-1)\cup[0,1).
\end{equation*}
If $\lambda$ satisfies the latter pair of conditions, then $z$ given by \eqref{eq:star_z} always belongs to $\sigma(D^0)$. On the other hand, if $\Re(\lambda)<0$, then $z$ given by \eqref{eq:star_z} is always away from $\sigma(D^0)$. Therefore, (ii) holds true.
\end{proof}

We see that $\sigma_p(D^{\Lambda_{A,B}})\cap\rho(D^0)$ may contain finitely many points, $\C^+$, $\C^-$ or even both these open half-planes. Moreover, it is possible to switch between these options by infinitesimal changes of $A$ and $B$. Take, for example, $A=\alpha I, B=\beta I$ with $\alpha,\beta\in\R$ and $I\equiv I_N$. Then, for $\alpha\neq 0$, $\sigma(\alpha I,0)=\emptyset$ and we are in the case (II); in fact, $D^{\Lambda_{\alpha I,0}}=D^0$. On the other hand, $\sigma(0,0)=\C$, $D^{\Lambda_{0,0}}=D^{max}$, and we are in the case (I). Finally, for $\alpha\neq 0$,  $\sigma(\alpha I,-\alpha I)=\{-1\}$. Therefore, we are in the case (III), and if $m>0$ then $z=0$ is the only eigenvalue of $\sigma_p(D^{\Lambda_{\alpha I,-\alpha I}})$ in $\rho(D^0)$.

The same abrupt spectral transitions were observed in \cite[Section 3.4]{HeTu_2022} for the one-dimensional non-self-adjoint Dirac operator with a non-self-adjoint point interaction, i.e., for the Dirac operator on the graph that consists of one vertex and two external edges -- one ingoing and one outgoing. Since, for simplicity, we deal with the outgoing edges in the current example, a direct comparison is not possible. One has to employ the parity transform, introduced in Section \ref{sec:parity} below, to compare the two settings.

We note that the conditions and results in (I)--(III) have a similar nature as those in the non-relativistic setting in \cite{HuKrSi_2015} for \emph{regular} and \emph{irregular} boundary conditions; see \cite[Equation (3.1)]{HuKrSi_2015}. In fact, boundary conditions with $A,B$ that satisfy $\rank(A\vert B)=N$ are regular exactly when the case (II) or (III) from Proposition \ref{prop:star_shaped} occurs and irregular exactly when (I) occurs.

\section{Symmetry transformations} \label{sec:symm}

By the principle of relativity, the behaviour of a physical system should not depend on the chosen inertial frame. In relativistic physics, the inertial frames are linked by Poincar\'{e} transformations of the space-time. Following \cite[Sect. 2]{thaller}, where the physical four-dimensional space-time is considered, we derive how some of these coordinate transformations correspond to the symmetry transformations of wave functions (more precisely rays in the projective space) that describe the relativistic particles on a line. The time-dependent Dirac equation in one spatial dimension reads as
\begin{equation} \label{eq:time_dep}
\iu\partial_t\psi=-\iu\pauli{1}\partial_x\psi+m\pauli{3}\psi,
\end{equation}
where $t$ stands for the time and the right-hand side is just the free Dirac operator $D$ on the line that is self-adjoint in $L^2(\R;\C^2)$ with $\Dom{D}=H^1(\R;\C^2)$. We will consider the so-called \emph{discrete  Lorentz} transformations, i.e. the space inversion $(\tilde t,\tilde x)=\mathcal{P}(t,x)=(t,-x)$ and the time reversal transformation $(\tilde t,\tilde x)=\mathcal{T}(t,x)=(-t,x)$. By the Wigner-Bargmann theorem \cite{Ba64}, the symmetry transformations are represented by unitary or antiunitary operators in $L^2(\R;\C^2)$. Therefore, it makes sense to assume that the symmetry transformations associated with $\mathcal{P}$ and $\mathcal{T}$ are of the form $(P\psi)(\tilde{t},\tilde{x})=U_P\psi(\mathcal{P}^{-1}(\tilde{t},\tilde{x}))$ and $(T\psi)(\tilde{t},\tilde{x})=U_T\psi(\mathcal{T}^{-1}(\tilde{t},\tilde{x}))$, respectively, where $U_i,\, i=P,T,$ is a unitary or antiunitary mapping in $\C^2$. 

It is straightforward to check that the equivalence ``$\psi$ solves \eqref{eq:time_dep} if and only of $P\psi$ solves the same equation in $(\tilde{t},\tilde{x})$-variables'' holds true if and only if 
$$U_P^{-1}\pauli{1} U_P=-\pauli{1}\quad \text{and} \quad U_P^{-1}\pauli{3} U_P=\pm\pauli{3},$$
where the $\pm$ sign corresponds to a unitary/antiunitary $U_P$. Note that when $m\neq0$ only for unitary $U_P$'s that satisfy these conditions, we have $\tilde{D} P=P D$, where $\tilde{D}$ is the free Dirac operator in $\tilde{x}$-variable. With an antiunitary $U_P$ one would get $\tilde{D} P=P D_{-m}$, where $D_{-m}$ stands for $D$ with $m$ replaced by $-m$.  Since, in the end, we will work with a time-independent Dirac equation, we prefer the situation when the free operator ``commutes'' with $P$ and we choose $U_P=\sigma_3$; this choice is unique up to a phase factor $\omega \in \mathbb{C}$ with $|\omega| = 1$ which has no physical significance. Similarly, in the case of $T$-symmetry, one would get the conditions
\begin{equation*}
U_T^{-1}\pauli{1} U_T=-\pauli{1}\quad \text{and} \quad U_T^{-1}\pauli{3} U_T=\mp\pauli{3}
\end{equation*}
under which $DT=\mp TD$, where the $\mp$ sign corresponds to a unitary/antiunitary choice of $U_T$. Since we again prefer  the situation when $D$ commutes with $T$ we choose $U_T=\pauli{3}\mathcal{C}$, where $\mathcal{C}$ stands for the complex conjugation. This choice is again unique up to a phase factor.

Beside symmetry transformations stemming from the invariance under Poincar\'{e} transformations, we will be also interested in the charge conjugation $C$ that is an internal symmetry that connects the negative energy subspace of particles (electrons) with the positive energy subspace of antiparticles (positrons). More concretely, it is an antiunitary transformation that obeys $CD(e)=-D(-e)C$, where $D(e)$ is the Dirac operator with the charge $e$ in the electromagnetic field (for the free operator $D(\pm e)=D$). Mimicking \cite[Sect. 1.4.6]{thaller} in our one-dimensional setting, we conclude that $C=U_C\mathcal{C}$, where the unitary matrix $U_C$ satisfies
$$\pauli{3}U_C=-U_C\pauli{3}\quad \text{and} \quad \pauli{1}U_C=U_C\pauli{1}.$$
Consequently, up to a phase factor, we are forced to choose $U_C=\pauli{1}$.

We summarize our findings in Table \ref{tab:symm}.
\begin{table}[h]
\begin{tabular}{|c|c|c|} 
\hline 
symmetry & origin & invariance \\ 
\hline 
$(P\psi)(x)=\pauli{3}\psi(-x)$ & space inversion & $\tilde{D}P=PD$ \\ 

$(T\psi)(x)=\pauli{3}\overline{\psi(x)}$ & time reversal & $DT=TD$ \\ 
$(C\psi)(x)=\pauli{1}\overline{\psi(x)}$ & charge/energy & $CD(e)=-D(-e)C$ \\ 
\hline 
\end{tabular} 
\caption{Symmetry transformations for the one-dimensional Dirac operator}
\label{tab:symm}
\end{table}
Note that up to a different representation of the Dirac algebra, the same transformations were derived in \cite{BoLuMa_2024}, where instead of $D$ a one-dimensional Dirac operator with a self-adjoint point interaction was considered. Whether the (anti)commutation properties with $P, T$ or $C$ are preserved strongly depends on the transmission conditions in the interaction point. Below, we will use the definitions of the symmetry transformations edge-wise and then we will find necessary and sufficient conditions under which a Dirac operator on a graph is $P,T$ or $C$-symmetric.
To this aim, every transformation will be firstly studied on a single edge, internal or external, and the obtained results will be then extended to the whole graph.

We emphasise that all results in this section hold true in the exactly same form both for local and non-local transmissions conditions.

\subsection{Change of orientation} \label{sec:parity}
In the case of the Dirac operator on a line, $P$ itself can be called parity or space inversion transformation. However, in the case of a graph we refrain from calling $P$ like that, as there is no global coordinate on the graph that can be inverted. We will rather think of $P$ as a \emph{change of orientation} on one or more (possibly all) edges. This is related to the question whether and how the spectral properties of the Dirac operator on a graph depend on a choice of orientation on the graph's edges. Keep in mind that our change of orientation is always accompanied by the action of $\pauli{3}$ on the components of wave functions. 

\subsubsection{Change of orientation on a single edge}
When it is possible, we treat internal and external edges at the same time and we do not specify their index unless there is a difference between them. In that case, we use $i$ for an internal and $e$ for an external edge.
The change of orientation arises from the change of coordinate on an edge. Suppose that the original orientation assigns to the edge the interval $I = \left( a, b \right)$ with the coordinate $x$. Then the opposite orientation assigns to the same edge the interval $\theother{I} = ( \theother{a}, \theother{b} )=( -b, -a )$ with the coordinate $\theother{x}=-x$. Note that for external edges $\theother{\orient}=-\orient$.
The associated Hilbert space also changes, because $P$ defined similarly as on the whole line, i.e.
\begin{equation} \label{eq:P_edge}
P\phi(\theother{x}) := \pauli{3} \phi(-\theother{x})\quad (\forall \phi \in L^2(I;\C^2), \forall \theother{x} \in \theother{I}),
\end{equation}
acts from $\HH := L^2(I;\C^2)$ to $\theother{\HH} := L^2(\theother{I};\C^2)$. 

Note that $P$ is a linear bijective isometry that is formally involutive and
\begin{equation} \label{eqDPcommute}
\theother{D}^{max}  P = P  D^{max},
\end{equation}
where  $D^{max}$ is the original maximal Dirac operator on $\HH$ and $\theother{D}^{max} := -\iu \partial_{\theother{x}} \pauli{1} + m \pauli{3}$ is the maximal Dirac operator on $\theother{\HH}$. Let $( \bma,\bmb,\boundspace )$ and $( \tobma, \tobmb, \boundspace )$, as defined in \eqref{eqDefBTIedge} for an internal edge with $\boundspace=\C^2$ and in \eqref{eqDefBTEedge} for an external edge with $\boundspace=\C$, be the boundary triples for $D^{max}$ and $\theother{D}^{max}$, respectively. The following results describe how their boundary mappings and the induced Krein $\gamma$-fields and the induced Krein $\Kreinfun$-functions are related.

\begin{proposition} \label{prop:edge_bm}
Define the linear involutive operator $V_P$ on $\boundspace$ as
\begin{equation}\label{eqVP}
\forall \omega \in \boundspace:
\qquad
V_P \omega := 
\begin{cases}
\pauli{1} \omega &\text{ for an internal edge, } \boundspace = \C^2,
\\
\omega &\text{ for an external edge, } \boundspace = \C.
\end{cases}
\end{equation}
Then
\begin{equation}\label{eqPbm}
\tobma  P = V_P  \bma 
\qquad \text{and} \qquad
\tobmb  P = V_P  \bmb.
\end{equation}
\end{proposition}
\begin{proof}
Take any $\phi \in H^1(I;\C^2)$. By the definitions of $\bm,\, \tobm$ and using $\theother{a}=-b,\, \theother{b}=-a$ we get for an internal edge
\begin{align*}
\tobma P\phi &= 
\vectorComp{ \left( P\phi\right)^1 (\theother{a}) \\ \left( P\phi\right)^1 (\theother{b}) } =
\vectorComp{ \phi^1 (-\theother{a}) \\ \phi^1 (-\theother{b}) } = 
\vectorComp{ \phi^1 (b) \\ \phi^1 (a) } = 
\pauli{1} \bma \phi
\end{align*}
and
\begin{align*}
\tobmb P\phi &= 
\vectorComp{ \iu \left( P\phi \right)^2 (\theother{a}) \\ -\iu \left( P\phi \right)^2 (\theother{b})} = 
\vectorComp{ -\iu \phi^2 (-\theother{a}) \\ + \iu \phi^2 (-\theother{b}) } = 
\vectorComp{ -\iu \phi^2 (b) \\ \iu \phi^2 (a)} = 
\pauli{1} \bmb \phi.
\end{align*} 
In a similar way, we get for an external edge $(a,+\infty)$
\begin{align*}
\tobma P\phi &=  (P\phi)^1 (\theother{b}) =
 (\pauli{3}\phi)^1 (-\theother{b}) = \phi^1 (a) = \bma \phi
\end{align*}
and
\begin{align*}
\tobmb P\phi &= -\iu(P\phi)^2 (\theother{b}) = 
-\iu  \left( \pauli{3}\phi \right)^2 (- \theother{b} ) =
\iu  \phi^2 ( a ) = 
\bmb \phi,
\end{align*}
and similarly for an external edge parametrized by $(-\infty,b)$.
\end{proof}

\begin{proposition}  \label{prop:edge_gamma} Let $D^0=D^{max} \upharpoonright {\Ker \bma}$, $\theother{D}^0=\theother{D}^{max} \upharpoonright \Ker{\tobma}$ and $\gamma(z)$ and $\theother{\gamma}(z)$ be the Krein $\gamma$-field induced by the triples $( \bma,\bmb,\boundspace )$ and $( \tobma,\tobmb,\boundspace )$, respectively. Then we have $\resolvent (\refoperator)=\resolvent (\theother{D}^0)$ and, for all $z \in \resolvent (\refoperator)$, 
\begin{equation}\label{eqPKreinG}
\theother{\gamma}(z)=P\gamma(z)V_P.
\end{equation}
\end{proposition}
\begin{proof}
First, since 
\begin{equation} \label{eqD_0Pcommute}
\theother{D}^0=P\refoperator P^{-1}
\end{equation}
and $P$ is a bijective isometry, we get $\resolvent (\refoperator)=\resolvent (\theother{D}^0)$. Similarly, as  $\theother{D}^{max}=PD^{max}P^{-1}$, we obtain
$$\theother{N}_z:=\Ker(\theother{D}^{max}-z)=\Ker(P(D^{max}-z)P^{-1})=P(N_z),$$
i.e. $P$ is a bijection between the deficiency spaces of $D^{max}$ and $\theother{D}^{max}$, which together with the first equality in \eqref{eqPbm} and the relation $V_P^{-1}=V_P$ implies that
$$\theother{\gamma}(z)=(\tobma\upharpoonright \theother{N}_z)^{-1}=(V_P\bma P^{-1}\upharpoonright P(N_z))^{-1}=(V_P(\bma\upharpoonright N_z) P^{-1})^{-1}=P\gamma(z)V_P.$$
\end{proof}

\begin{proposition} \label{prop:edge_Krein} Let $\Kreinfun$ and $\theother{\Kreinfun}$ be the Krein $\Kreinfun$-functions induced by the triples $( \bma,\bmb,\boundspace )$ and $( \tobma,\tobmb,\boundspace )$, respectively.  Then, for every $z \in \resolvent (\refoperator)$, we have
\begin{equation}\label{eqPKreinQ}
\theother{\Kreinfun}(z) = V_P  \Kreinfun (z)  V_P = \Kreinfun(z).
\end{equation}
\end{proposition}
\begin{proof}
Using  the second equality in \eqref{eqPbm} together with \eqref{eqPKreinG} we see that
\begin{equation*}
\theother{\Kreinfun}(z) = \tobmb  \theother{\gamma}(z) = \tobmb  P  \gamma(z)  V_P = V_P  \bmb  \gamma(z)  V_P = V_P  \Kreinfun(z)  V_P.
\end{equation*}
The second equality in \eqref{eqPKreinQ} follows from the explicit form of $V_P$ and $\Kreinfun(z)$, see \eqref{eqVP}, \eqref{Weyl_function_internal_edge_z}, \eqref{Weyl_function_internal_edge_m}, and~\eqref{eqKreinQEedge}. 
\end{proof}

\subsubsection{Change of orientation on a graph}

Let $\graph\equiv (\edges,\vertices,\gb)$ be an oriented graph, $\edges_{-}$ a subset of its edges $\edges$, and $\theother{\graph}\equiv(\edges,\vertices,\theother{\gb})$  another oriented graph that is constructed from $\graph$ by redefining its boundary map as follows,
\begin{equation*}
\theother{\gb} j=
\begin{cases}
\gb j =(\gb_- j, \gb_+ j),& j\in\edges\setminus\edges_{-},\\
(\gb_+ j, \gb_-j), & j\in\edges_{-},
\end{cases}
\end{equation*}
i.e. for the edges in $\edges_-$ the roles of incoming and outgoing edges with respect to any vertex are swapped. Moreover, we adopt a convention that if in the graph $\graph$ an edge $j$ is assigned an interval $I_j=(a,b)$ then in the graph $\theother{\graph}$ the same edge is assigned an interval
\begin{equation*}
\theother{I}_j=(\theother{a},\theother{b})=
\begin{cases}
I_j=(a,b), & j\in\edges\setminus\edges_{-},\\
(-b,-a), & j\in\edges_{-}.
\end{cases}
\end{equation*}
Note that this choice is compatible with our previous convention that the initial and the final endpoint of an edge  correspond to the left and the right endpoint, respectively, of the assigned interval. Moreover, the maximal operator on $\theother{\graph}$ is unitarily equivalent to the one on $\graph$, as the operations on $\tilde{I}_j$ and $I_j$ can be related via a shift from $(-b,-a)$ to $(a,b)$. 
Finally, if $\HH = \bigoplus_{j\in\edges} \HH_j$ with $\HH_j = L^2(I_j;\C^2)$ is the Hilbert space associated with $\graph$ then  $\theother{\HH} = \bigoplus_{j\in\edges} \theother{\HH}_j$ with $\theother{\HH}_j = L^2(\theother{I}_j;\C^2)$ is the Hilbert space associated with $\theother{\graph}$.

Below, we will use the following notation
\begin{equation*}
\edges_{+} := \edges \setminus \edges_{-},
\qquad
\intes_{\pm} := \intes \cap \edges_{\pm},
\qquad
\extes_{\pm} := \extes \cap \edges_{\pm}.
\end{equation*}
If $\orient(e)$ describes the orientation of an external edge $e\in\extes$ in $\graph$ then the orientation of $e$ in $\theother{\graph}$ is given by
\begin{equation*}
\theother{\orient}(e) =
\begin{cases} \orient(e), & \text{if } e \in \extes_+,\\
- \orient(e), & \text{if } e \in \extes_-\,.
\end{cases}
\end{equation*} 
Moreover, in both cases, we have $\theother{\gb}_{\theother{\orient}(e)}e = \gb_{\orient(e)}e \in \vertices$, where $\theother{\gb}$ stands for the graph boundary map of $\theother{\graph}$.

Recall that the maximal Dirac operator on the graph $\graph$ is the direct sum of maximal Dirac operators on its edges, $D^{max} = \bigoplus_{j\in\edges} D^{max}_{j}$. The maximal operator on the graph $\theother{\graph}$ is $\theother{D}^{max} = \bigoplus_{j\in\edges} \theother{D}^{max}_{j}$, where $\theother{D}^{max}_{j}$ is just the maximal Dirac operator in $\theother{\HH}_j$. Similarly, we introduce the reference operator $D^0$ and $\theother{D}^0$ in $\HH$ and $\theother{\HH}$, respectively. Next, we will construct a bijective isometry $P_{\edges_{-}}:\HH\to\theother{\HH}$ that acts as the change of orientation operator \eqref{eq:P_edge} on the edges in $\edges_-$ and as the identity on the remaining edges. Namely, we define
\begin{equation*}
(P_j\phi)(\theother{x}) := \pauli{3} \phi(-\theother{x})\quad (\forall\phi \in\HH_j=L^2(I_j;\C^2), \forall \theother{x} \in \theother{I}_j)
\end{equation*}
and 
\begin{equation} \label{eq:P_I-_def}
P_{\edges_{-}} := \bigoplus_{j\in\edges} P_{\edges_{-},j}\quad \text{with} \quad
P_{\edges_{-},j}:=
\begin{cases}
I_{\HH_j}, & j\in\edges_+,\\
P_j, & j\in\edges_-\, .
\end{cases}
\end{equation}
Using \eqref{eqDPcommute} and \eqref{eqD_0Pcommute} together with the fact that all operators $D^{max},\, \theother{D}^{max},\, D^0,\, \theother{D}^0,$ and $P_{\edges_-}$ decompose to direct sums, we infer that $P_{\edges_-}$ leaves the maximal and the reference operator invariant in the sense that
\begin{equation} \label{eq:D_maxD_0P_comm}
\theother{D}^{max} P_{\edges_{-}} = P_{\edges_{-}} D^{max}\quad \text{and} \quad \theother{D}^{0}  P_{\edges_{-}} = P_{\edges_{-}} D^{0}.
\end{equation}
In particular, the latter implies that 
\begin{equation} \label{eq:res_refoperators}
\rho(\theother{D}^0)=\rho(\refoperator).
\end{equation}

Next, we will investigate how the boundary triples and the induced Krein $\gamma$-fields and the induced Krein $\Kreinfun$-functions for $D^{max}$ and $\theother{D}^{max}$ are related.  

\begin{proposition} \label{prop:gamma_Q_P_trafo}
Let $(\bma, \bmb, \boundspace)$  and $(\tobma,\tobmb,\boundspace)$, constructed as in Proposition \ref{prop:totalBT}, be the boundary triples for $D^{max}$ and $\theother{D}^{max}$, respectively.  Define the linear involutive operator $V_{P_{\edges_{-}}}$ on the common boundary space $\boundspace$ as
\begin{equation}\label{eqVPedges-}
V_{P,\edges_{-}} := \bigoplus_{j \in \edges} V_{P,\edges_{-},j}
\quad \text{with} \quad
V_{P,\edges_{-},j}
:=
\begin{cases}
I_\C, & j \in \extes, \\
\pauli{0}, & j \in \intes_+, \\
\pauli{1}, & j \in \intes_-\,.
\end{cases}
\end{equation}
Then
\begin{equation}\label{eqPbmGraph}
\tobma P_{\edges_{-}} = V_{P,\edges_{-}}  \bma 
\quad \text{and} \quad
\tobmb  P_{\edges_{-}} = V_{P,\edges_{-}}  \bmb.
\end{equation}
Furthermore, for $z\in\rho(\refoperator)$, the induced Krein $\gamma$-fields and the induced Krein $\Kreinfun$-functions obey
\begin{equation} \label{eqPKreinQgraph}
\theother{\gamma}(z)=P_{\edges_{-}} \gamma(z)V_{P,\edges_{-}}  
\quad \text{and} \quad \theother{\Kreinfun}(z) = V_{P,\edges_{-}}  \Kreinfun (z)  V_{P,\edges_{-}} = \Kreinfun(z).
\end{equation}
\end{proposition}
\begin{proof}
All operators in \eqref{eqPbmGraph} and in \eqref{eqPKreinQgraph} decompose with respect to the direct sum decompositions of $\HH$, $\theother{\HH}$, and $\boundspace$. 
Hence, it is sufficient to check all the equalities edge-wise. To prove \eqref{eqPbmGraph} we use Proposition \ref{prop:edge_bm} on the edges from $\edges_-$. On the remaining edges, the equalities follow trivially. 

Recall \eqref{eq:res_refoperators} and note that $\resolvent(\refoperator) = \bigcap_{j\in\edges} \resolvent(\refoperator_{j})$. Therefore, all objects in \eqref{eqPKreinQgraph} are well defined when $z\in\rho(\refoperator)$ and, in the same vein as above, we may use Propositions \ref{prop:edge_gamma} and \ref{prop:edge_Krein} to show \eqref{eqPKreinQgraph}.
\end{proof}

Now, we will investigate how the isomorphisms $W$ and $\theother{W}$ between the boundary space $\boundspace$ and the vertex space $\vertspace$ are related. Note that not only $\boundspace$ but also the vertex space $\vertspace$ with its structure $\bigoplus_{v\in\vertices} \vertspace_{v}$ is identical for both graphs $\graph$ and $\theother{\graph}$. The vertex-edge numbering $\nu_{\vertspace}$ does not change either, as it only depends on the orders of the vertices and of the edges. These sets are the same, so the order can be assumed to be the same as well. However, the edge-vertex numberings are generally different, since for $i\in\intes_{-}$ the roles of the first (left) vertex and the second (right) vertex are interchanged. 

Denote the edge-vertex numberings for $\graph$ and $\theother{\graph}$ by $\nu_{\boundspace}$ and $\theother{\nu}_{\boundspace}$, respectively.  Recall  that $W$ can be thought of as a permutation matrix for the permutation $\nu_{\vertspace}^{-1} \nu_{\boundspace}$. Similarly, $\tilde{W}$ will correspond to $\nu_{\vertspace}^{-1}  \theother{\nu}_{\boundspace}$. Then we may write
\begin{equation}\label{eqWotherIsWS}
\tilde{W} = WS,
\end{equation}
where  $S$ is the matrix corresponding to the permutation $\nu_{\boundspace}^{-1} \theother{\nu}_{{\boundspace}}$, i.e.
\begin{equation} \label{eq:Sdef}
S_{k,l} =
\begin{cases}
1, & \nu_{\boundspace}(k) = \theother{\nu}_{\boundspace}(l),\\
0, & \text{otherwise}.
\end{cases}
\end{equation}
In fact, $S$ viewed as an operator in the boundary space $\boundspace$ acts in the same way as $V_{P,\edges_{-}}$.

\begin{proposition} 
Let $S$ given by \eqref{eq:Sdef} be understood as an operator in $\boundspace$. Then $S = V_{P,\edges_{-}}$, where $V_{P,\edges_{-}}$ is defined in \eqref{eqVPedges-}.
\end{proposition}
\begin{proof}
Since for both edge-vertex numberings the order of edges is identical and the numbers of the incident vertices are common as well, $S$ decomposes to a direct sum with respect the decomposition $\boundspace=\bigoplus_{j\in\edges}\boundspace_j$ and $\nu_{\boundspace}$ and $\theother{\nu}_{\boundspace}$ only act differently when their first component maps to $\intes_-$. If the latter occurs, then the vertices incident with $j\in\intes_-$ that are numbered in one way by $\nu_{\boundspace}$ are numbered the other way around by $\theother{\nu}_{\boundspace}$, i.e. $S$ acts as $\pauli{1}$ on the corresponding subspace $\boundspace_j$. The claim follows by a direct comparison with $\eqref{eqVPedges-}$.
\end{proof}

Finally, we are ready to compare the Dirac operator $D^{\Lambda_{W}}$ on $\graph$ defined by \eqref{DLambdaDomain}, i.e. $D^{\Lambda_W} = D^{max} \upharpoonright \Dom D^{\Lambda_W}$, where
\begin{equation*}
\Dom D^{\Lambda_W} = \set{ \Phi \in \widetilde{H}^1(\graph;\C^2)}{(W \bma \Phi,W \bmb \Phi) \in \Lambda},
\end{equation*}
to the Dirac operator $\theother{D}^{\Lambda_{\tilde{W}}}$ defined similarly on 
$\theother{\graph}$ by $\theother{D}^{\Lambda_{\tilde{W}}} = \theother{D}^{max} \upharpoonright \Dom \theother{D}^{\Lambda_{\tilde{W}}}$ with 
\begin{equation*}
\Dom \theother{D}^{\Lambda_{\tilde{W}}} = \set{ \theother{\Phi} \in \widetilde{H}^1(\theother{\graph};\C^2)}{(\tilde{W} {\tobma} \theother{\Phi},\tilde{W} {\tobmb} \theother{\Phi}) \in \Lambda}.
\end{equation*}

\begin{theorem}\label{thmPDLambda}
Let $\Lambda$ be a linear relation in the common vertex space $\vertspace$ of the graphs $\graph$ and $\theother{\graph}$, $D^{\Lambda_W}$ and $\theother{D}^{\Lambda_{\tilde{W}}}$ be as above and $P_{\edges_{-}}$ as in \eqref{eq:P_I-_def}. Then
\begin{equation*}
\theother{D}^{\Lambda_{\tilde{W}}}P_{\edges_{-}}=P_{\edges_{-}}D^{\Lambda_W}.  
\end{equation*}
In particular, $\theother{D}^{\Lambda_{\tilde{W}}}$ and $D^{\Lambda_W}$ are unitarily equivalent.
\end{theorem}
\begin{proof}
Taking into account the first equality in \eqref{eq:D_maxD_0P_comm} and the fact that $D^{\Lambda_W}$ and $\theother{D}^{\Lambda_{\tilde{W}}}$ are restrictions of $D^{max}$ and $\theother{D}^{max}$, respectively, it remains to show that $P_{\edges_{-}} (\Dom D^{\Lambda_W} ) = \Dom \theother{D}^{\Lambda_{\tilde{W}}}$.
Using \eqref{eqPbmGraph}, \eqref{eqWotherIsWS} and $S=V_{P,\edges_{-}}=V_{P,\edges_{-}}^{-1}$, we get, for $k\in\lbrace 1,2 \rbrace$ and $\Phi \in \widetilde{H}^1(\graph;\C^2)$,
\begin{equation*}
\tilde{W} \tobm^{k} P_{\edges_{-}} \Phi = W S V_{P,\edges_{-}} \bm^{k} \Phi = W \bm^{k} \Phi.
\end{equation*}
Therefore, for every $\Phi \in \widetilde{H}^1(\graph;\C^2)$, we have 
$$(W \bma \Phi,W \bmb \Phi) = (\tilde{W} {\tobma} P_{\edges_{-}} \Phi,\tilde{W} {\tobmb} P_{\edges_{-}} \Phi).$$
Taking the definitions of $D^{\Lambda_W}$ and $\theother{D}^{\Lambda_{\tilde{W}}}$ into account, we conclude that the latter equality implies $P_{\edges_{-}} (\Dom D^{\Lambda_W} ) = \Dom \theother{D}^{\Lambda_{\tilde{W}}}$.
\end{proof}

\begin{remark}
  Since $\theother{D}^{\Lambda_{\tilde{W}}}$ and $D^{\Lambda_W}$ are unitarily equivalent, their spectra of all types coincide and their resolvents are also linked via the connecting unitary map $P_{\edges_{-}}$. Similar results can also be seen with the help of the boundary triple $(\tobma,\tobmb,\boundspace)$. Indeed, with the induced Krein $\gamma$-field $\theother{\gamma}$ and the induced Krein $\Kreinfun$-function $\theother{\Kreinfun}$ from~\eqref{eqPKreinQgraph}, the spectrum of $\theother{D}^{\Lambda_{\tilde{W}}}$ can be described with the Birman Schwinger principle from \cite[Theorem~2.6.2]{BeHaSn_2020}, whose  variant for the relations of the form \eqref{eq:def_Lambda_AB} is given in Theorem~\ref{theorem_Birman_Schwinger}. Taking the form of $\theother{\Kreinfun}$ into account, one finds for $z \in \rho(D^0) = \rho(\theother{D}^0)$ that
  \begin{equation*}
   z \in \spectrum_p(\theother{D}^{\Lambda_{\tilde{W}}}) \, \Leftrightarrow \, 0 \in \sigma(\Lambda - \tilde{W}\tilde{\Kreinfun}(z) \tilde{W}^{-1}) = \sigma(\Lambda - {W}{\Kreinfun}(z) {W}^{-1}) \, \Leftrightarrow \, z \in \spectrum_p(D^{\Lambda_W}),
  \end{equation*} 
  and that $\rho(\theother{D}^0) \cap (\spectrum_r(\theother{D}^{\Lambda_{\tilde{W}}}) \cup \spectrum_c(\theother{D}^{\Lambda_{\tilde{W}}})) = \emptyset$,
  yielding $\sigma(\theother{D}^{\Lambda_{\tilde{W}}}) \cap \rho(\theother{D}^0) = \sigma(D^{\Lambda_W}) \cap \rho(D^0)$. Nevertheless, note that the transformed Birman Schwinger principle does not simplify the analysis. Similar statements are true for the Krein resolvent formula from  \cite[Theorem~2.6.1]{BeHaSn_2020} or Theorem~\ref{theorem_Birman_Schwinger}(iii) for relations of the form \eqref{eq:def_Lambda_AB}.
\end{remark}

\subsection{Time reversal transformation} \label{sec:Treversal}

\subsubsection{Time reversal transformation on a single edges}

If a single edge is parametrized by $x\in I = \left( a,b \right)$ then the Hilbert space associated with the edge is just $\HH = L^2(I;\C^2)$ and the \emph{time reversal transformation} is the antilinear map defined by
\begin{equation*}
T\phi := \pauli{3} \cc{\phi}\quad (\forall \phi \in \HH).
\end{equation*}
It is immediate to check that
\begin{equation}\label{eqDTcommute}
D^{max}  T = T  D^{max}.
\end{equation}

\begin{proposition}
Let $( \bma,\bmb,\boundspace )$ be the boundary triple for $D^{max}$ as defined in \eqref{eqDefBTIedge} and \eqref{eqDefBTEedge} for an internal and an external edge, respectively. Define an antilinear involutive isometry $V_T$ on $\boundspace$ as a complex conjugation of the vector entries,
\begin{equation*} 
\forall \omega \in \boundspace:
\qquad
V_T \omega := \cc{\omega}. 
\end{equation*}
Then
\begin{equation}\label{eqTbm}
\bma  T = V_T  \bma 
\qquad \text{and} \qquad
\bmb  T = V_T  \bmb.
\end{equation}
\end{proposition}

\begin{proof}
Take any $\phi \in \Dom(D^{max})=H^1(I;\C^2)$. Then for an internal edge it holds
\begin{align*}
\bma T\phi &= \vectorComp{ (T\phi)^1 (a) \\ (T\phi)^1 (b) } = \vectorComp{ \cc{\phi^1(a)} \\ \cc{\phi^1(b)} } = \cc{ \bma\phi } = V_T \bma \phi
\end{align*}
and
\begin{align*}
\bmb T\phi &= \vectorComp{ \iu (T\phi)^2 (a) \\ -\iu (T\phi)^2 (b) } = \vectorComp{ \iu ( \cc{-\phi^2(a)} ) \\ -\iu ( \cc{-\phi^2(b)} ) } = \cc{ \bmb\phi } = V_T \bmb \phi.
\end{align*}
In a similar way, we get for an external edge $(a,+\infty)$ 
\begin{align*}
\bma T\phi &=  (T\phi)^1 (a)  =  \cc{ \phi^1(a) }  = V_T \bma \phi
\end{align*}
and
\begin{align*}
\bmb T\phi &=   \iu (T\phi)^2 (a)  =  -\iu \cc{ \phi^2(a) }  = V_T \bmb \phi,
\end{align*}
and similarly for an edge parametrized by $(-\infty,b)$.
\end{proof}

\begin{proposition} \label{prop:TKrein}
For $z \in \resolvent (\refoperator)$, the Krein $\gamma$-field induced by the triple $(\bma, \bmb, \boundspace)$ transforms as follows under the action of $T$,
\begin{equation}\label{eqTKreinG}
T \gamma(z) = {\gamma}(\cc{z}) V_T.
\end{equation}
\end{proposition}
\begin{proof}
First of all, note that $z\in \resolvent(\refoperator)$ if and only if $\cc{z} \in \resolvent (\refoperator)$, due to self-adjointness of $\refoperator$. Therefore, $\gamma(z)$ and $\gamma(\cc{z})$ exist simultaneously.
Now, $\gamma(z)$ takes the boundary values $\omega \in \boundspace$ to $\phi\in\Dom(D^{max})$ such that $(D^{max}-z)\phi=0$ and $\bma\phi=\omega$. Using \eqref{eqDTcommute} and \eqref{eqTbm} we get
$$0=T(D^{max}-z)\phi=(D^{max}-\cc{z})T\phi \quad \text{and} \quad \bma T\phi=V_T\bma \phi=V_T\omega,$$
which implies that $T\gamma(z)\omega=T\phi=\gamma(\cc{z})V_T\omega$, i.e. \eqref{eqTKreinG}.
\end{proof}

\begin{proposition} \label{prop:TKreinQ}
For $z \in \resolvent (\refoperator)$, the Krein $\Kreinfun$-function induced by the triple $(\bma, \bmb, \boundspace)$ is transformed in the following way,
\begin{equation*}
{\Kreinfun}(\cc{z}) = V_T  \Kreinfun (z)  V_T.
\end{equation*}
\end{proposition}
\begin{proof}
Again, note that $\Kreinfun(z)$ and $\Kreinfun(\cc{z})$ exist simultaneously. Using \eqref{eqTKreinG}, \eqref{eqTbm}, and $V_T^{-1}=V_T$, we obtain
\begin{equation*}
\Kreinfun(\cc{z}) = 
\bmb \gamma(\cc{z}) =
\bmb T \gamma(z) V_T =
V_T \bmb \gamma(z) V_T =
V_T \Kreinfun(z) V_T.
\end{equation*}
\end{proof}

\subsubsection{Time reversal transformation on a graph}

The time reversal transformation on an oriented graph $\graph$ is defined as the following antilinear involutive isometry $T:\HH=L^2(\graph;\C^2)\to\HH$,
\begin{equation*}
T := \bigoplus_{j\in\edges} T_j \quad \text{with} \quad  T_j \phi_j:= \pauli{3} \cc{\phi_j} \quad (\forall \phi_j \in \HH_j).
\end{equation*}
Since the maximal Dirac operator on $\graph$ decomposes into a direct sum with respect to the direct sum decomposition of $\HH$, we immediately get
\begin{equation} \label{eq:D_maxT_comm}
D^{max}  T = T  D^{max},
\end{equation}
due to \eqref{eqDTcommute}.

First, we will check how the boundary mappings and the induced Krein $\gamma$-field and the induced Krein $\Kreinfun$-function behave under the action of $T$. We will use the notation $V_T$ for an antilinear involutive isometry on the boundary or vertex space acting as the complex conjugation on every vector component, i.e.
$$V_T\omega:=\cc{\omega}\quad (\forall\omega\in\boundspace,\,\vertspace).$$
\begin{proposition}
Let $( \bma, \bmb, \boundspace )$ be the boundary triple for the maximal Dirac operator $D^{max}$ on a graph introduced in Proposition \ref{prop:totalBT}. Then the following holds
\begin{equation}\label{eqTbmGraph}
\bma  T = V_T  \bma 
\qquad \text{and} \qquad
\bmb  T = V_T  \bmb.
\end{equation}
\end{proposition}
\begin{proof}
Since
\begin{equation*}
V_T = \bigoplus_{j\in\edges} V_{T,j} \quad \text{with} \quad
 V_{T,j} \omega_j = \cc{\omega_j} \quad (\forall \omega_j\in \boundspace_j),
\end{equation*}
every map in \eqref{eqTbmGraph} is a direct sum with respect to the considered decompositions of $\boundspace$ and $\HH$. Hence, it is enough to use \eqref{eqTbm} on every edge.
\end{proof}

\begin{corollary}\label{cor:WbmTisVTWbm}
Let $W: \boundspace \to \vertspace$ be the isomorphism from the boundary space onto the vertex space and $\bma,\, \bmb,$ and $V_T$ be as above. Then
\begin{equation*}
W \bma T = V_T W \bma 
\quad \text{and} \quad
W \bmb T = V_T W \bmb.
\end{equation*}
\end{corollary}

\begin{proof}
Using \eqref{eqTbmGraph} and the fact that the entries of $W$ are real-valued (either $0$ or $1$), we get, for $k\in\lbrace 1,2 \rbrace$, 
\begin{equation*}
W \bm^{k} T  = W V_T \bm^{k}  = V_T W \bm^{k}.
\end{equation*}
\end{proof}

\begin{proposition}\label{propTKreinGgraph}
For every $z \in \resolvent (\refoperator)$, we have
\begin{equation*}
T  \gamma(z) = \gamma(\cc{z})  V_T \quad \text{and} \quad \Kreinfun(\cc{z}) = V_T  \Kreinfun(z)  V_T.
\end{equation*}
\end{proposition}
\begin{proof}
One can use a similar argument as in the proof of Proposition \ref{prop:gamma_Q_P_trafo} and apply Propositions \ref{prop:TKrein} and \ref{prop:TKreinQ} edge-wise.
\end{proof}

\begin{theorem}
Let $\Lambda$ be a linear relation in the vertex space $\vertspace$ and $\cc{\Lambda}$ be the linear relation consisting of the complex-conjugated elements of $\Lambda$, i.e. 
$$\cc{\Lambda} := \set{(\cc{f},\cc{f'})}{(f,f')\in\Lambda}.$$
Then it holds 
\begin{equation*}
T D^{\Lambda_W} =  D^{\cc{\Lambda}_W}T.
\end{equation*}
In particular, $D^{\Lambda_W}$ and $D^{\cc{\Lambda}_W}$ are (antilinearly) similar to each other.
\end{theorem}
\begin{proof}
Due to \eqref{eq:D_maxT_comm}, it remains to check that $T(\Dom{D^{\Lambda_W}})=\Dom{D^{\cc{\Lambda}_W}}$. Take $\Phi\in\Dom{D^{max}}$. 
Then $\Phi\in\Dom{D^{{\Lambda_W}}}$ if and only if $\left(W\bma\Phi, W\bmb\Phi \right) \in {\Lambda}$, which is equivalent to $\left(V_T  W\bma\Phi, V_T W\bmb\Phi \right) \in \cc{\Lambda}$. Since, by  Corollary \ref{cor:WbmTisVTWbm}, 
\begin{equation*}
  \left( V_T W\bma\Phi, V_T W\bmb\Phi \right) = \left( W\bma T\Phi, W\bmb T\Phi \right),
\end{equation*}
the latter occurs if and only if $T\Phi\in\Dom{D^{\cc{\Lambda}_W}}$. 
\end{proof}

\begin{corollary} \label{corollary_spectrum_time_reversal}
The operator $D^{\Lambda_W}$ is \emph{$T$-symmetric} if and only if $\Lambda=\cc{\Lambda}$. For $\Lambda_W=\Lambda_{A,B}$, this occurs if and only if there exists a linear bijection $X$ on $\vertspace$ such that $\cc{A}=XA$ and $\cc{B}=XB$, cf. Proposition \ref{prop:equivRep}. 

Furthermore, due to (antilinear) similarity, we have
\begin{equation*}
\lambda \in \spectrum_i (D^{{\Lambda_W}}) 
\Leftrightarrow 
\cc{\lambda} \in \spectrum_i (D^{\cc{\Lambda}_W}),\quad i\in\{p,c,r,d,ess\}.
\end{equation*}
In particular, if $D^{\Lambda_W}$ is $T$-symmetric, then its spectrum is symmetric with respect to the complex conjugation.
\end{corollary}

\begin{remark} \label{remark_T}
  Parts of the results from Corollary~\ref{corollary_spectrum_time_reversal} can also be seen with the help of the boundary triple $(\bma,\bmb,\boundspace)$. Indeed, with  Proposition~\ref{propTKreinGgraph}, the Birman Schwinger principle from  \cite[Theorem~2.6.2]{BeHaSn_2020}, whose  variant for the relations of the form \eqref{eq:def_Lambda_AB} is given in Theorem~\ref{theorem_Birman_Schwinger}, and $V_T \Lambda_W V_T = \overline{\Lambda}_W$, one finds for $z\in\rho(D^0)$  that
  \begin{equation*}
    \begin{split}
      z \in \spectrum_p(D^{\Lambda_W}) \quad &\Leftrightarrow \quad 0 \in \sigma(\Lambda_W - \Kreinfun(z)) \\
      & \Leftrightarrow \quad  0 \in \sigma(V_T (\Lambda_W - \Kreinfun(z)) V_T) = \sigma(\overline{\Lambda}_W - \Kreinfun(\overline{z})) \\
      &\Leftrightarrow \quad \overline{z} \in \spectrum_p(D^{ \overline{\Lambda}_W}).
    \end{split}
  \end{equation*}
  Nevertheless, note again that the transformed Birman Schwinger principle does not simplify the analysis. Similar statements are true for the Krein resolvent formula from \cite[Theorem~2.6.1]{BeHaSn_2020} or Theorem~\ref{theorem_Birman_Schwinger}(iii) for relations of the form \eqref{eq:def_Lambda_AB}.
\end{remark}

\subsection{Charge conjugation transformation} \label{sec:Ctrafo}

\subsubsection{Single edges}
Let $\HH = L^2(I;\C^2)$ be again the Hilbert space associated with a single edge. The \emph{charge conjugation transformation} is defined as the following antilinear involutive isometry in $\HH$,
\begin{equation*}
C\phi :=\pauli{1} \cc{\phi} \quad (\forall \phi \in \HH).
\end{equation*}
One easily checks that $C$ anticommutes with the maximal Dirac operator $D^{max}$ on $\HH$, i.e.,
\begin{equation}\label{eqDCanticommute}
D^{max}  C = -C  D^{max}.
\end{equation}

The transformation $C$ on $\HH$ has again its counterpart on the boundary space $\boundspace$. However, note that the roles of the boundary maps $(\bma,\bmb)$ are now intertwined.

\begin{proposition} \label{prop:C_bm_single}
Let $( \bma,\bmb,\boundspace )$ be the boundary triple for $D^{max}$ as defined in \eqref{eqDefBTIedge} and \eqref{eqDefBTEedge} for an internal and an external edge, respectively, and $V_C$ be the following antilinear involutive isometry on $\boundspace$,
\begin{equation*}
V_C \omega := 
\begin{cases}
 +\iu \pauli{3} \cc{\omega} & \text{ for an internal edge} \quad (\forall\omega\in\boundspace = \C^2),
\\
 -\iu \orient \cc{\omega} & \text{ for an external edge} \quad (\forall\omega\in\boundspace = \C),
\end{cases}
\end{equation*}
where $\rho$ is the orientation of the external edge, i.e., $\rho=-1$ for $I=(a,+\infty)$  and $\rho=1$ for $I=(-\infty,b)$.
Then there holds
\begin{equation}\label{eqCbm}
\bma  C = V_C  \bmb
\qquad \text{and} \qquad
\bmb  C = V_C  \bma.
\end{equation}
\end{proposition}
\begin{proof}
Take any $\phi \in\Dom(D^{max})=H^1(I;\C^2)$. Then for an internal edge it holds
\begin{align*}
\bma C\phi 
&= \vectorComp{ (\pauli{1} \cc{\phi})^1 (a) \\ (\pauli{1} \cc{\phi})^1 (b) } 
= \vectorComp{ \cc{\phi^2 (a)} \\ \cc{\phi^2 (b)} } 
= \begin{pmatrix} \iu & 0 \\ 0 & -\iu \end{pmatrix} \vectorComp{ \cc{\iu \phi^2 (a) } \\ \cc{-\iu \phi^2 (b) } }
= V_C \bmb \phi
\end{align*}
and
\begin{align*}
\bmb C\phi 
&= \vectorComp{ \iu (\pauli{1} \cc{\phi})^2 (a) \\ -\iu (\pauli{1} \cc{\phi})^2 (b) } 
= \vectorComp{ \iu \cc{\phi^1 (a)} \\ -\iu \cc{\phi^1 (b)} } 
= \begin{pmatrix} \iu & 0 \\ 0 & -\iu \end{pmatrix} \vectorComp{ \cc{ \phi^1 (a) } \\ \cc{ \phi^1 (b) } }
= V_C \bma \phi.
\end{align*}
In a similar way, for an external edge $I=(a,+\infty)$ we get
\begin{align*}
\bma C\phi 
&=  \left( \pauli{1} \cc{\phi} \right)^1 (a) 
=  \cc{\phi^2 (a) } 
= \iu  \cc{(\iu  \phi^2 (a) ) }
= \iu  \cc{\bmb \phi}
\end{align*}
and
\begin{align*}
\bmb C\phi 
&=  \iu  \left( \pauli{1} \cc{\phi} \right)^2 (a)  
= \iu  \cc{\phi^1 (a) } 
= \iu  \cc{\bma \phi},
\end{align*}
and similarly when $I=(-\infty,b)$.
\end{proof}


\begin{proposition} \label{prop:C_KreinQ}
For $z \in \resolvent (\refoperator)$ in the case of an external edge and for $z \in \resolvent (\refoperator)\setminus \lbrace m \rbrace$ in the case of an internal edge, we have
\begin{equation}\label{eqCKreinQ}
{\Kreinfun}(-\cc{z}) = V_C  \Kreinfun(z)^{-1}  V_C.
\end{equation}
\end{proposition}
\begin{proof}
First of all, since $\sigma(\refoperator)\subset\R$ and $\sigma(\refoperator)=-\sigma(\refoperator)$ except for the case of a finite edge and $m\neq 0$ (when $-m\in\sigma(\refoperator)$ but $m\notin\sigma(\refoperator)$; cf. Propositions~\ref{proposition_D_0_internal_edge} and~\ref{proposition_D_0_external_edge}), the both sides of \eqref{eqCKreinQ} make sense simultaneously. Next, employing \eqref{eqDCanticommute}, it is straightforward to check that $C$ is a bijection between the deficiency spaces of $D^{\max}$ with $z\in\C$ and $-\cc{z}$, i.e., 
\begin{equation} \label{eqCDefectSub}
C(N_z) = N_{-\cc{z}}.
\end{equation}
Now, restricting \eqref{eqCbm} to $N_z$, it follows from  \eqref{eqCDefectSub} that for $k \in \lbrace 1,2 \rbrace$, $\left( \bm^{k} \upharpoonright N_{-\cc{z}} \right)  C = V_C  \left( \bm^{3-k} \upharpoonright N_{z} \right)$. Note that $\bmb \upharpoonright N_{z} = V_{C}^{-1}  (\bma \upharpoonright N_{-\cc{z}})  C$ is invertible if and only if $-\cc{z} \in \resolvent(\refoperator)$. For such $z$, we get
\begin{multline*}
\Kreinfun(-\cc{z})
=
\left( \bmb \upharpoonright N_{-\cc{z}} \right)C C^{-1}	 \left( \bma \upharpoonright N_{-\cc{z}} \right)^{-1}
=
\left( (\bmb \upharpoonright N_{-\cc{z}})  C \right) 	\left( (\bma \upharpoonright N_{-\cc{z}})  C \right)^{-1}
=
\\
=
\left( V_C  \bma \upharpoonright N_{z} \right) 	\left( V_C  \bmb \upharpoonright N_{z} \right)^{-1}
=
V_C \Kreinfun(z)^{-1} V_C^{-1}=V_C \Kreinfun(z)^{-1} V_C.
\end{multline*}
\end{proof}

\subsubsection{Charge conjugation transformation on a graph}
The charge conjugation transformation on an oriented graph $\graph$ is defined as the following antilinear involutive isometry $C:\HH\to\HH$,
\begin{equation*}
C := \bigoplus_{j\in\edges} C_j \quad \text{with} \quad C_j \phi_j := \pauli{1} \cc{\phi_j} \quad (\forall \phi_j \in \HH_j).
\end{equation*}
Due to \eqref{eqDCanticommute} and the fact that both $C$ and the maximal Dirac operator on $\graph$ decompose with respect to the same direct sum decomposition of $\HH$, we get
\begin{equation} \label{eq:DmaxCanticomm}
D^{max} C = - C  D^{max}.
\end{equation}
For a similar reason, we obtain, employing Propositions \ref{prop:C_bm_single} and \ref{prop:C_KreinQ}, the following result. 

\begin{proposition}
Let $( \bma, \bmb, \boundspace )$ be the boundary triple for the maximal Dirac operator $D^{max}$ on $\graph$ constructed in Proposition \ref{prop:totalBT} and  $V_C$ be the following antilinear involutive isometry on $\boundspace$,
\begin{equation*}
V_C := \bigoplus_{j\in\edges} V_{C,j} \quad \text{with} \quad
 V_{C,j} \omega_j := 
\begin{cases}
\iu \pauli{3} \cc{\omega_j}, & j \in \intes, \\
-\iu \rho(j) \cc{\omega_j}, & j \in \extes,
\end{cases}
\quad (\forall \omega_j\in \boundspace_j).
\end{equation*}
Then there holds
\begin{equation}\label{eqCbmGraph}
\bma  C = V_C \bmb 
\quad \text{and} \quad
\bmb  C = V_C \bma.
\end{equation}
Moreover, for $z\in\rho(\refoperator)\setminus\{m\}$, we have
\begin{equation*}
\Kreinfun(-\cc{z}) = V_C  \Kreinfun(z)^{-1}  V_C.
\end{equation*}
\end{proposition}

\begin{theorem} \label{theorem_C}
Let $\Lambda$ be a linear relation in the vertex space $\vertspace$ and 
$$\Lambda_C := \set{(V_C W^{-1}{f'}, V_C W^{-1}{f})}{(f,f')\in\Lambda},$$
where $W: \boundspace \to \vertspace$ stands for the isomorphism from the boundary space onto the vertex space.
Then it holds 
\begin{equation*}
D^{\Lambda_C}C =  -CD^{\Lambda_W}.
\end{equation*}
\end{theorem}

\begin{proof}
Due to \eqref{eq:DmaxCanticomm}, it remains to check that $C\Dom{D^{\Lambda_W}}=\Dom{D^{\Lambda_C}}$.
Let $\Phi\in\Dom{D^{max}}=\widetilde{H}^1(\graph;\C^2)$. Then, using \eqref{eqCbmGraph}, we get
\begin{equation*}
\left( \bma C\Phi,  \bmb C\Phi \right)
= \left(  V_C \bmb \Phi,  V_C \bma \Phi \right) 
=  \left((V_C W^{-1}) W \bmb \Phi,(V_C W^{-1}) W \bma \Phi \right).
\end{equation*}
Hence,  $\Phi\in\Dom{D^{\Lambda_{W}}}$, i.e.  $\left( W \bma \Phi , W \bmb \Phi \right)\in\Lambda$, if and only if $\left(\bma C\Phi , \bmb C\Phi \right)\in\Lambda_C$, i.e., $C\Phi\in\Dom{D^{\Lambda_C}}$.
\end{proof}

\begin{remark} 
Note that $V_C$ acts as a composition of the complex conjugation in all components with the following linear mapping
\begin{equation*}
\tilde{V}_C := \bigoplus_{j\in\edges} \tilde{V}_{C,j}, \quad \text{where} \quad
 \tilde{V}_{C,j} \omega_j := 
\begin{cases}
\iu \pauli{3} \omega_j, & j \in \intes, \\
-\iu \rho(j)  \omega_j, & j \in \extes,
\end{cases}
\quad (\forall \omega_j\in \boundspace_j).
\end{equation*}
Since, viewed as a matrix, $W$ is a permutation matrix and $\tilde{V}_C$ is diagonal, it follows that $W\tilde{V}_CW^{-1}$ is also diagonal and its diagonal elements are given just by the permutation $W$ of the diagonal elements of $\tilde{V}_C$.

\end{remark}

\begin{corollary} \label{corollary_C}
The operator $D^{\Lambda_W}$ is \emph{$C$-symmetric} if and only if $\Lambda_W=\Lambda_C$. For $\Lambda_W=\Lambda_{A,B}$, this occurs if and only if there exists a linear bijection $X$ on $\vertspace$ such that $\cc{B}W\tilde{V}_C W^{-1}=XA$ and $\cc{A}W\tilde{V}_CW^{-1}=XB$, cf. Proposition \ref{prop:equivRep}. 

Furthermore, due to (antilinear) anti-similarity, we have
\begin{equation*}
\lambda \in \spectrum_i (D^{\Lambda_W}) 
\Leftrightarrow 
-\cc{\lambda} \in \spectrum_i (D^{\Lambda_C}),\quad i\in\{p,c,r,d,ess\}.
\end{equation*}
In particular, if $D^{\Lambda_W}$ is $C$-symmetric, then its spectrum is symmetric with respect to the reflection over the imaginary axis.
\end{corollary}

\begin{remark}
  We note that the relation of the spectra of $D^{\Lambda_W}$ and $D^{\Lambda_C}$ in Corollary~\ref{corollary_C}, that follows from their (antilinear) anti-similarity in Theorem~\ref{theorem_C}, can also be shown, in a similar fashion as in Remark~\ref{remark_T}, with the help of the Birman Schwinger principle and Proposition~\ref{prop:C_KreinQ}. We leave the details to the reader.
\end{remark}

\begin{remark} \label{rem:C_symm_star}
The $C$-symmetry condition simplifies considerably for the star shaped graph with external outgoing edges only, that was introduced in Section \ref{sec:star_shaped}, because then $W=I_N$ and $\tilde{V}_C=\iu I_N$. Therefore $D^{\Lambda_{A,B}}$ on such a graph is $C$-symmetric if and only if there exists a linear bijection $X$ on $\vertspace$ such that $\cc{B}=XA$ and $\cc{A}=XB$.
\end{remark}

\begin{appendix}
\section{Proof of Theorem~\ref{theorem_Birman_Schwinger}} \label{appendix_relations}

To show Theorem~\ref{theorem_Birman_Schwinger}, we show first some auxiliary results on linear relations, which are mostly true also for infinite-dimensional spaces $\boundspace$.

\begin{lemma} \label{lem:invertibility}
Let $\Lambda$ be a closed relation in $\boundspace$ ($\dim(\boundspace)\leq+\infty$) of the form \eqref{def_Lambda} and $R$ be a bounded operator in $\boundspace$. Then $(\Lambda-R)^{-1}$ exists as a bounded operator in $\boundspace$ if and only if $\Ker(A-BR)=\{0\}$ and $\Ran{B}\subset\Ran(A-BR)$.
\end{lemma}

\begin{proof}
Since $\Lambda-R$ is a closed relation, $(\Lambda-R)^{-1}$ exists as a bounded operator in $\boundspace$ if and only if $\Ker(\Lambda-R)=\{0\}$ and $\Ran(\Lambda-R)=\boundspace$. Using
\begin{equation*}
\Lambda-R=\{(f,f'-Rf)\,|\, Af=Bf'\},
\end{equation*}
we infer that $f\in \Ker(\Lambda-R)$ if and only if there exists $f'$ such that $f'=Rf$ and $Af=Bf'$, which is equivalent to the condition $Af=BRf$, i.e. $f\in\Ker(A-BR)$. Next, $g\in\Ran(\Lambda-R)$ if and only if there exist $(f,f')\in\boundspace \oplus \boundspace$ such that $g=f'-Rf$ and $Af=Bf'$. Expressing $f'=g+Rf$, we see that this happens if and only if there exists $f\in\boundspace$ such that $(A-BR)f=Bg$, i.e. $Bg\in\Ran(A-BR)$.

Therefore,  $\Ker(\Lambda-R)=\{0\}$ if and only if $\Ker(A-BR)=\{0\}$ and $\Ran(\Lambda-R)=\boundspace$ if and only if $\Ran{B}\subset\Ran(A-BR)$.
\end{proof}

If $\dim \boundspace < +\infty$, then a linear operator in $\boundspace$ is surjective if and only if it is injective and hence, Lemma~\ref{lem:invertibility} simplifies in the following way.

\begin{corollary} \label{corollary_finite_dimensional}
If $\dim(\boundspace)<+\infty$ then $(\Lambda-R)^{-1}$ exists as a bounded operator in $\boundspace$ if and only if $\Ker(A-BR)=\{0\}$, i.e. $0\notin\sigma(A-BR)$.
\end{corollary}

With the help of Lemma~\ref{lem:invertibility} we can prove now the following variant of Krein's resolvent formula for $S^\Lambda$.

\begin{theorem} \label{theorem_krein}
Let $(\bma, \bmb, \boundspace)$ be a boundary triple for $S^*$ (with $\dim{\boundspace}\leq+\infty$), let $S^0 = S^* \upharpoonright \Ker \bma$, and let $\gamma$ and $\Kreinfun$ be the induced Krein $\gamma$-field and the induced Krein $\Kreinfun$-function, respectively. Moreover, let $\Lambda$ be  a linear relation in $\boundspace$ of the form~\eqref{def_Lambda} and let $S^\Lambda$ be given by~\eqref{dom_S_Lambda}. Then, for every $z\in\rho(S^0)$, there holds that $z\in\rho(S^\Lambda)$ if and only if $\Ker(A-B\Kreinfun(z))=\{0\}$ and $\Ran{B}\subset\Ran(A-B\Kreinfun(z))$. If this is the case then 
\begin{equation*} 
(S^\Lambda - z)^{-1} = (S^0 - z)^{-1} + \gamma(z) \bigl( A - B \Kreinfun(z) \bigr)^{-1} B \gamma(\overline{z})^*.
\end{equation*}
\end{theorem}

\begin{proof}
The first statement is just a consequence of \cite[Theorem~2.6.2 (iv)]{BeHaSn_2020}, \cite[Lemma 1.2.4 (i)]{BeHaSn_2020} and Lemma \ref{lem:invertibility}. The resolvent formula follows from \cite[Theorem~2.6.2 (i)]{BeHaSn_2020} and the observation that
\begin{equation*}
(\Lambda-\Kreinfun(z))=\{(f,g)\,|\, (A-B\Kreinfun(z))f=Bg\}=\{((A-B\Kreinfun(z))^{-1}Bg,g)\,|\, g\in\boundspace\}.
\end{equation*}
\end{proof}

\begin{proof}[Proof of Theorem~\ref{theorem_Birman_Schwinger}]
The claims in~(i) and (iii) follow from Theorem~\ref{theorem_krein}, Corollary~\ref{corollary_finite_dimensional}, and \cite[Theorem~2.6.2 (i)]{BeHaSn_2020}. For (ii), one concludes first from (i) taking the assumption $\rank(A|B) < \dim \boundspace$ into account that $\C \setminus \R \subset \sigma_p(S^\Lambda)$. Since $\sigma(S^\Lambda)$ is closed, this implies that $\sigma(S^\Lambda) = \C$.
\end{proof}

\end{appendix}

\subsection*{Acknowledgements}
The authors thank the Ministry of Education, Youth and Sports of the Czech Republic, the BMBWF, and the Austrian Agency for International Cooperation in Education and Research (OeAD) for support within the project 8J23AT025 and CZ 16/2023, respectively.

\end{document}